\documentclass[footinbib,showpacs,preprint,amsmath,amssymb,11pt]{article}                                          
\usepackage[a4paper]{geometry}                                              \usepackage[outercaption]{sidecap}                                     
\usepackage{graphicx}       \usepackage{xcolor}
\usepackage{bm}
\usepackage{amsmath, amsthm, amssymb}
\usepackage{braket}        
\usepackage[colorlinks=true,linkcolor=blue,citecolor=blue,urlcolor=blue]{hyperref}

\renewcommand\Im{\operatorname{Im}}
\renewcommand\Re{\operatorname{Re}}

\definecolor{orange}{rgb}{1,0.5,0}
\usepackage{scrextend}
\newcommand{\bek}{\begin{eqnarray}}
\newcommand{\ek}{\end{eqnarray}}
\usepackage[caption=false,margin=0pt]{subfig}
\newcommand{\VEC}[1]{\mathbf{#1}}
\newcommand{\ud}{\mathrm{d}}
\newcommand{\iu}{\mathrm{i}}

\begin{document}
\parskip 1ex

\begin{center}
\section*{ \includegraphics[scale=0.15]{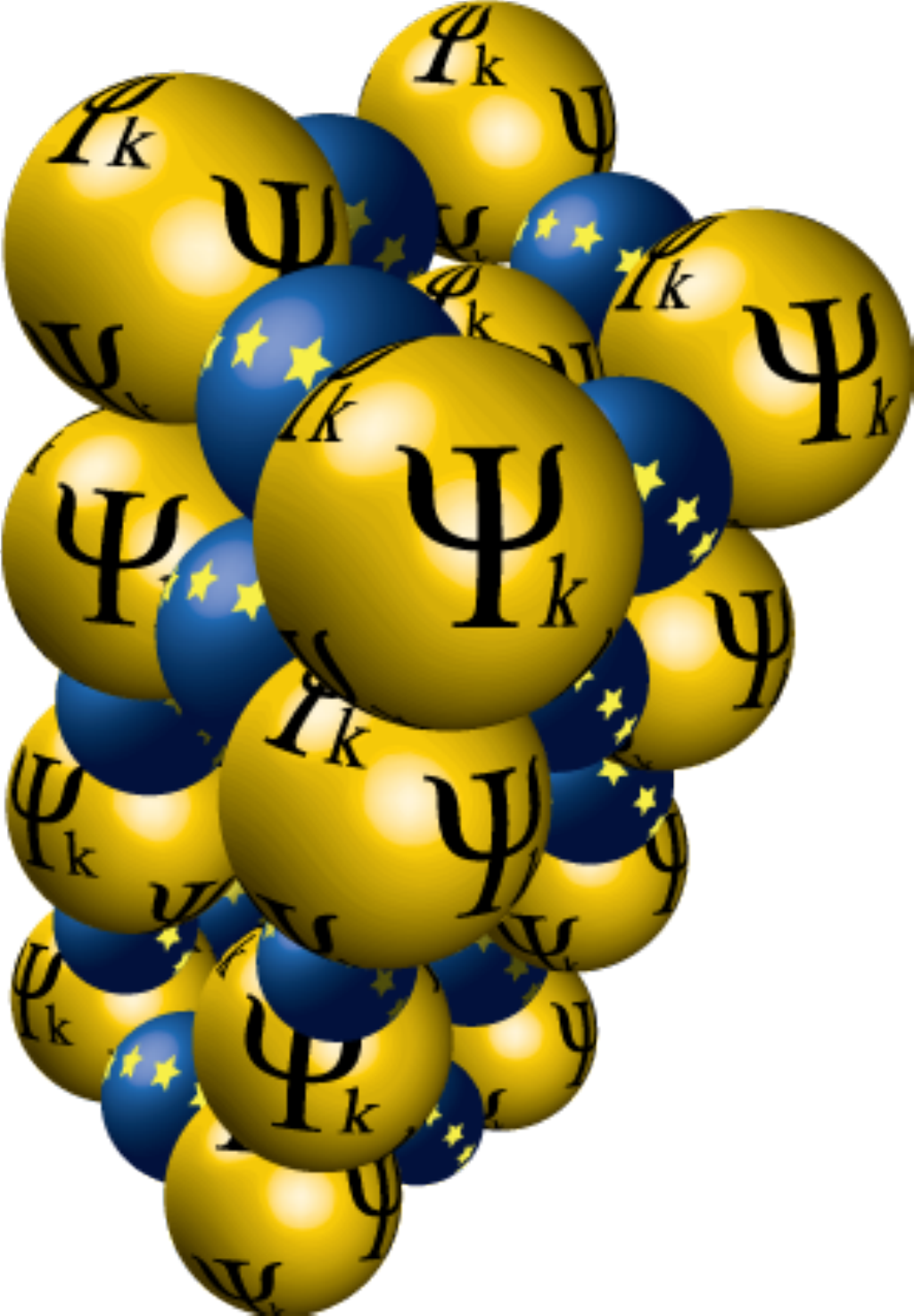}
           \Large $\Psi_k$ Scientific Highlight of the Month }
\end{center}
{\bf No. 140 \hfill March 2018}
\vspace{0.1cm}
\rule{16.0cm}{1mm}
\vspace{2mm}
\setcounter{section}{0}
\setcounter{figure}{0}

\begin{center}

{\large \bf Spin-fluctuation and spin-relaxation effects of single adatoms from first principles}

Julen Iba\~{n}ez-Azpiroz$^{2}$, Manuel dos Santos Dias$^{1}$,  Stefan Bl\"ugel$^{1}$, 
Samir Lounis$^{1}$\\

{\it $^{1}$ Peter Gr\"unberg Institut and Institute for Advanced Simulation, Forschungszentrum
J\"ulich \& JARA, 52425 J\"ulich, Germany}\\
{\it $^{2}$ Centro de F{\'i}sica de Materiales, 
Universidad del Pa{\'i}s Vasco, 20018 San Sebasti{\'a}n, Spain}
\end{center}



%

\begin{abstract}

Single adatoms offer an exceptional playground for studying magnetism and its associated dynamics at the atomic scale.
Here we review recent results on single adatoms deposited on metallic substrates, based on time-dependent density functional theory.
First we analyze quantum zero-point spin-fluctuations (ZPSF) as calculated from the fluctuation-dissipation theorem, and show how they affect the magnetic stability by modifying the magnetic anisotropy energy.
We also assess the impact of ZPSF in the limit of small hybridization to the substrate characteristic of semi-insulating substrates, connecting to recent experimental investigations where magnetic stability of a single adatom was achieved for the first time. 
Secondly, we inspect further the dynamics of single adatoms by considering the longitudinal and transverse spin-relaxation processes, whose time-scales are analyzed and related to the underlying electronic structure of both the adatom and the substrate.
Thirdly, we analyze spin-fluctuation modes of paramagnetic adatoms, i.e., adatoms where the Stoner criterion for magnetism is almost fulfilled. 
Interestingly, such modes can develop well-defined peaks in the meV range, their main characteristics being determined by two fundamental electronic properties, namely the Stoner parameter and the density of states at the Fermi level.
Furthermore, simulated inelastic scanning tunneling spectroscopy curves reveal that these spin-fluctuation modes can be triggered by tunneling electrons, opening up potential applications also for paramagnetic adatoms.
Lastly, an overview of the outstanding issues and future directions is given.

\end{abstract}


Single adatoms deposited on surfaces have become a prominent  playground where theory and experiment can explore hand-in-hand a large variety of physical phenomena ranging from spin-excitations~\cite{heinrich_single-atom_2004,Hirjibehedin2006,hirjibehedin_large_2007,Balashov2009,Lorente2009,Rossier2009,lounis_dynamical_2010,Khajetoorians2011,Khajetoorians2013,Donati2013,ternes_spin_2015,khajetoorians_tailoring_2016,guimaraes_2017,hermenau_long_2018} to magnetic exchange interactions~\cite{Hirjibehedin2006,Meier2008,zhou_strength_2010,Strozecka2011,Khajetoorians2012,oberg_control_2014,yan_control_2015}, quantum spin decoherence~\cite{bryant_controlled_2015,baumann_electron_2015,delgado_spin_2017}, interplay between magnetism and superconductivity~\cite{Ji2008,Braunecker2013,Heinrich2013,Vazifeh2013,nadj-perge_observation_2014,Ruby2015,Ruby2016}, or the Kondo effect~\cite{Madhavan1998,Neel2008,Otte2008,Otte2009,ternes_spectroscopic_2009,Delgado2014,vonBergmann2015}, among many others. 
Virtually all these effects arise from the intricate interplay between the degrees of freedom of the adatom --- charge, spin or orbital angular momentum --- and the electron and phonon bath of the substrate, a subject of heavy and ongoing investigation. 
Noteworthily, magnetism plays a central role in fuelling the interest for these tiny objects, given that they represent the ultimate limit in the context of bit miniaturization in data storage devices. 
Therefore, understanding the above mentioned phenomena is of capital importance both for fundamental physics and for potential technological applications.
In this context, the field has witnessed an enormous push forward after the advent of spin-polarized scanning tunnelling microscopy (STM) and inelastic electron tunnelling spectroscopy (IETS).
These experimental techniques, occasionally used in combination with X-ray magnetic circular dichroism (XMCD)~\cite{donati_magnetic_2016,Brune2009} and electron paramagnetic resonance (EPR)~\cite{baumann_electron_2015}, allow  to monitor the dynamical regime by, e.g., measuring atomic spin-excitations~\cite{heinrich_single-atom_2004,hirjibehedin_large_2007,otte_role_2008,Khajetoorians2011} and quasiparticle interferences~\cite{Strozecka2011}, accessing spin-relaxation and decoherence times~\cite{loth_measurement_2010,baumann_electron_2015,oberg_control_2014} and even resolving highly dynamical processes like the reading and writing of magnetic information~\cite{natterer_reading_2017}.

From the theoretical point of view, early simulations based on density functional theory (DFT)~\cite{oswald_giant_1986,wildberger_magnetic_1995,lang_local_1994} served for boosting and creating a huge enthusiasm in the field by predicting gigantic local magnetic moments of diverse transition-metal (TM) adatoms, including 4\textit{d} and 5\textit{d} elements that are nominally nonmagnetic in bulk.
In addition, the understanding and interpretation of features present in IETS experiments, such as the characteristic steps in the  conductance spectrum~\cite{hirjibehedin_spin_2006}, greatly benefited from studies based on quantum spin models, e.g., by showing the importance of inelastic excitation channels involved in the process (see, e.g., 
Refs.~\cite{lorente_efficient_2009,Rossier2009,Persson2009,fransson_spin_2009,sothmann_nonequilibrium_2010,Gauyacq2012}).
In the last few years, the development of a real-space spin-dynamics technique within time-dependent DFT (TDDFT)~\cite{lounis_dynamical_2010,lounis_theory_2011} has proven to be a powerful tool for characterizing the central dynamical quantity, namely the spin-excitation spectrum, which contains information about the exchange of energy and angular momentum with the substrate electrons that form the environment of the magnetic adatoms~\cite{lounis_dynamical_2010,lounis_theory_2011}.
Importantly, a quantitative analysis of the role played by the spin-orbit coupling (SOC) in determining the resonance frequency of spin-excitations and their reaction to magnetic fields has been successfully given within this approach~\cite{dias_relativistic_2015}.
Going one step beyond, the use of many-body perturbation theory (MBPT) in connection with TDDFT has also made possible to establish a direct quantitative comparison to the $\mathrm{d}I/\mathrm{d}V$ (differential tunneling conductance) curves measured in IETS experiments~\cite{Schweflinghaus2014}.

One of the most fundamental requirements for a successful usage of single adatoms for technological purposes is the stability of its magnetic moment.
This, in turn, depends crucially on the so-called magnetic anisotropy energy (MAE), an energy barrier generated by the SOC that protects and stabilizes  the direction of the local magnetic moment against possible fluctuations of the spin, e.g., of thermal origin.   
Noteworthily, XMCD and IETS assert that several 3\textit{d} transition metal (TM) adatoms can possess a substantial MAE of few meV: examples include Fe and Co on Pt(111)~\cite{gambardella_giant_2003,Khajetoorians2013}, Fe on Cu(111)~\cite{Khajetoorians2011} as well as on Cu$_{2}$Ni/Cu(111)~\cite{Bryant2013} and CuNi~\cite{hirjibehedin_large_2007}, and Co on MgO(100)~\cite{rau_reaching_2014}, among others.
Strikingly, however, these large MAEs do not directly translate into an stable magnetic moment, given that the above mentioned systems behave as paramagnetic entities when measured by means of STM (see, e.g.,Refs.~\cite{meier_revealing_2008,khajetoorians_atom-by-atom_2012,zhou_strength_2010}), thus implying the existence of a mechanism that destroys the magnetic stability locally.
Going one step further, the case of 4\textit{d} and 5\textit{d} adatoms is even more striking, given that they have so far exhibited no clear magnetic signal even when subjected to the static magnetic field of an XMCD experiment, in notorious disagreement with theoretical predictions~\cite{honolka_absence_2007}. 
Up to date, the only single adatom that has experimentally exhibited magnetic remanence is Ho on MgO/Ag(100)~\cite{donati_magnetic_2016,natterer_reading_2017}, i.e., a 4\textit{f} rare-earth element with a very large local magnetic moment ($\sim10$ $\mu_{\mathrm{B}}$) deposited on a thin insulating layer.

In view of this rich scenario, a central question arises: what is the mechanism leading to the apparent instability of the magnetic moment of an adatom?
Given that temperature effects are not the principal cause of the magnetic instability~\cite{meier_revealing_2008}, one must turn to other destabilizing effects.
Recent efforts to address this far-reaching issue have been mainly based on symmetry arguments applied to model Heisenberg Hamiltonians, whereby the switching probability of the magnetic moment is calculated by means of a master equation (see e.g. Refs.~\cite{miyamachi_stabilizing_2013,hubner_symmetry_2014,khajetoorians_current-driven_2013,donati_magnetic_2016}).
These arguments apply to a well-localized magnetic moment of either integer or half-integer spin, which is justified in the weak coupling regime to the substrate.
If the coupling is strong, as it is generally the case for metallic substrates, the orbitals responsible for the magnetic moment of an adatom hybridize with the surface conduction electrons, giving rise to an itinerant electron picture with fractional local magnetic moments (see, e.g., Refs.~\cite{Khajetoorians2011,lang_local_1994,lounis_dynamical_2010,lounis_theory_2011,dias_relativistic_2015}).
In the latter case, the proper description of the itinerant character is needed for addressing the subject of magnetic stability.

The present review focuses on the aforementioned strongly-hybridized metallic systems.
In this context, zero-point spin-fluctuations (ZPSF) become a key dynamical aspect which are intimately connected to spin-excitations through the so-called fluctuation-dissipation theorem~\cite{Callen1951}.
ZPSF are quantum fluctuations present even at absolute zero temperature that can crucially affect the magnetic properties of itinerant electron magnets. 
The importance of spin-fluctuations was first realized in the early studies of weak bulk ferromagnets~\cite{moriya-book,takahashi,lonzarich_effect_1985,shimizu_itinerant_1981} and has subsequently been verified in the context of DFT and TDDFT too~\cite{aguayo_why_2004,ortenzi_accounting_2012}.
In this review, we consider the concept of ZPSF in single adatoms and quantify their impact on the magnetic stability of the series of 3\textit{d} and 4\textit{d} TM adatoms deposited on metallic substrates.
We show that ZPSF are  of the order of the local magnetic moment itself, an extremely large value that has deep effects on the MAE, which can be reduced by more than $50\%$ with respect to the static value calculated by standard DFT.
We further characterize the three fundamental ingredients that determine the magnitude of the ZPSF, namely the ($i$) local magnetic moment, ($ii$) SOC and ($iii$) electron-hole Stoner excitations.
Based on our findings, we develop a simple diagram where the ZPSF of an arbitrary adatom are classified according to these ingredients, offering practical guidelines for achieving magnetic properties which are robust against fluctuations.

Apart from protecting the magnetic stability, in order to achieve the ultimate goal of a technologically useful single adatom it is mandatory to master the dynamical properties associated to its spin.
For example, fast spin-dynamics can  be useful when the goal is to transfer magnetic information from or to the adatom, while slow spin-dynamics are desirable if the aim is to store magnetic information.
What are the typical time scales involved, and how does the environment affect them?
Interestingly, the lifetime of an atomic spin-excitation can be experimentally accessed from the width of the step observed in IETS $\text{d}I/\text{d}V$ measurements, being typically of the order of picoseconds~\cite{Khajetoorians2011,rau_reaching_2014,hirjibehedin_large_2007,loth_njp}.
Throughout this review, we quantitatively connect this lifetime to electronic properties of adatoms and their hybridization with the substrate, showing that the single-particle spin-flip excitation channel is the main driving mechanism behind it. 
In tandem, recent experimental developments yielded another way to measure spin-relaxation times using an all-electrical pump-probe technique~\cite{loth_measurement_2010}, which has been so far applied to adatoms deposited on insulating substrates~\cite{rau_reaching_2014,baumann_electron_2015,paul_control_2017}.
Based on the \textit{ab initio} perspective, here we show that, in order to successfully perform such experiments in metallic systems, a measuring scheme based on ultrafast techniques able to record relaxation times of the order of femtoseconds would be needed.

Although magnetic adatoms have by far attracted most of the interest in this field, nominally non-magnetic adatoms can also exhibit interesting fingerprints of magnetism and thus  be potentially useful for technological purposes. 
This is the last main topic addressed in the present review, where we study what are the requirements for observing well-defined features in the spin-excitation spectrum of this type of adatoms, {i.e.}, paramagnetic spin-excitations (PSE). 
These are the analogous of so-called paramagnons first proposed by Doniach in 1967~\cite{doniach_theory_1967} and first measured in bulk Pd nearly 50 years later by Doubble and co-workers~\cite{Doubble2010} (see also Ref.~\cite{staunton_spin_2000} for recent calculations). 
Upon reducing the dimensionality of the system, here we show that PSE can be strongly enhanced due to the modified interplay between the two fundamental electronic properties involved, namely the Stoner exchange interaction and the adatom's density of states (DOS) at the Fermi level.
In addition, the \textit{ab initio} analysis reveals that PSE are highly sensitive to externally applied magnetic fields and, furthermore, can exhibit a singular enhancement when the field approaches a critical regime.
Finally, we assess the impact of PSE on the $\mathrm{d}I/\mathrm{d}V$ signal as measured in IETS experiments, identifying clear signatures of magnetic response that allow to distinguish these type of excitations from, e.g., phonons.

The review is organized as follows. 
In Sec.~\ref{sec:theory}, we sketch the theoretical formalism used to calculate the quantities presented throughout the review.
In Sec.~\ref{sec:gs} we begin by presenting the ground state properties of a set of 3\textit{d} and 4\textit{d} TM adatoms that will be the main focus of the review.
In Sec.~\ref{sec:zpsf} the concept of ZPSF is presented as well as an analysis of its magnitude and its effect on inducing magnetic instability via the renormalization of the MAE barrier. 
In Section \ref{subsec:relax}, we focus on discussing atomic time scales for the relaxation time of the spin. 
Finally, in Sec.~\ref{sec:param} we illustrate the concept of paramagnetic spin-fluctuation modes and possible means of probing it experimentally.
Conclusions and outlook are given in Sec. \ref{sec:conclusions}, while appendices \ref{appendix:Bloch-long} and \ref{appendix:LLG} contain derivations on the Bloch and Landau-Lifshitz-Gilbert models, respectively.

\section{Theoretical approach}
\label{sec:theory}

In this section we settle the theoretical background used throughout the review. 
For this, let us consider the linear response of a ferromagnetic
system to an externally applied time-dependent perturbation,
\bek\label{eq:general-response}
  \delta {\textbf{m}}(\textbf{r};t) = \int\!\mathrm{d}\textbf{r}'\!\int\!\mathrm{d}t'\;
\boldsymbol{\chi}(\textbf{r},\textbf{r}';t-t')\,\delta {\textbf{V}}(\textbf{r}';t').
\ek
Above, $\delta{\textbf{m}}=(\delta m_{x},\delta m_{y},\delta m_{z},\delta n)$ and 
$\delta{\textbf{V}}=(\delta B_{x},\delta B_{y},\delta B_{z},\delta V)$, with 
$\delta m_{i}$ and $\delta B_{i}$ respectively 
the components of the vector spin density and external magnetic field, while
$\delta n$ and $\delta V$ are the charge density and external scalar potential, respectively. 
Working in frequency space and defining atomic-like quantities by integrating 
out the spatial dependence over atomic sites~\cite{lounis_theory_2011}, the above expression
takes the simplified form
\bek\label{eq:general-response-w}
\delta {\textbf{m}}(\omega)=
\boldsymbol{\chi}(\omega)\,\delta {\textbf{V}}(\omega).
\ek
For a single magnetic atom, the quantity $\boldsymbol{\chi}$ in the above equations is a 4$\times$4
tensor that couples in general all components of the spin and charge responses
with each other.
If SOC is weak, however,
the full response decouples into a longitudinal and transverse
part~\cite{book_vignale}. 
This approximation is justified for the systems investigated in this review since the 
off-diagonal sectors of the susceptibility tensor are small in comparison to the diagonal ones.
Then, assuming that the perturbation is purely of
magnetic origin (i.e. $\delta V=0$), 
the change of the spin magnetic moment length
is described by
\bek\label{eq:dmz-w}
\delta m_{z}(\omega)=
\chi_{\parallel}(\omega)\,\delta B_{z}(\omega).
\ek
Above, $\chi_{\parallel}(\omega)$ 
denotes the longitudinal spin-susceptibility.
This quantity is determined by excitations between electrons with same spin state,
given that it involves the Pauli matrix $\sigma_{z}$ that is
diagonal in spin basis~\cite{PhysRevLett.119.017203}.
On the other hand, the change of the transverse spin components 
can be compactly described using the circular combinations $m_{\pm}=m_{x}\pm \mathrm{i}\,m_{y}$ and 
$B_{\pm}=B_{x}\pm \mathrm{i}\,B_{y}$, yielding for the $+$ component
\bek\label{eq:dm+_w}
\delta m_{+}(\omega)=
\chi_{\pm}(\omega)\,\delta B_{+}(\omega).
\ek
Above, $\chi_{\pm}(\omega)$ denotes the transverse spin-susceptibility which, contrary to 
$\chi_{\parallel}(\omega)$, is determined by transitions that flip the spin state of the electrons
due to the transverse Pauli spin matrices involved, 
which are off-diagonal in spin space~\cite{lounis_dynamical_2010,lounis_theory_2011,dias_relativistic_2015}. 
We note that the full transverse response is given by
\bek
\chi_{\perp}(\omega) \equiv 
2\big( \chi_{\pm}(\omega)+\chi_{\mp}(\omega)\big).
\ek

The spin-susceptibility is closely connected to the ZPSF
via the so-called fluctuation-dissipation theorem~\cite{Callen1951}. This theorem
relates the variance of the spin-fluctuations along a given direction, 
$\xi^{2}_{i}$, 
to the imaginary part of the corresponding  spin-susceptibility, 
$\Im \chi_{i}(\omega)$, 
\begin{equation}
\label{eq:fd-theorem}
\xi_{i}^{2}= -\dfrac{1}{\pi} \int_{0}^{\infty}\!\!\!\! \mathrm{d}\omega\,\Im \chi_{i}(\omega),
\end{equation}
with $i={\parallel},\perp$.

The calculation of the ZPSF requires therefore a realistic and accurate description 
of the spin-susceptibility of the adatom.
This can be achieved by employing the time-dependent generalization of DFT, 
namely TDDFT~\cite{PhysRevLett.52.997,tddft-2}.
The general expression for the adatom's enhanced spin-susceptibility 
within the TDDFT framework is given by~\cite{lounis_dynamical_2010,lounis_theory_2011,dias_relativistic_2015,PhysRevB.91.104420,Schweflinghaus2014,ibanez-azpiroz_zero-point_2016}
\begin{equation}
\label{eq:susc-long}
\chi_{i}(\omega)=\dfrac{\chi^{\mathrm{KS}}_{i}(\omega)}{1-U_{i}\, \chi_{i}^{\mathrm{KS}}(\omega)},
\end{equation}
with $i={\parallel},\pm,\mp$. 
Above, $\chi^{\mathrm{KS}}_{i}(\omega)$ is the bare KS spin-susceptibility, i.e.
the standard response function in linear response theory. 
For the analysis provided in this review, we calculate the different components
of  $\chi^{\mathrm{KS}}_{i}(\omega)$ using the Korringa-Kohn-Rostoker Green function (KKRGF)
method employing  
the atomic sphere approximation 
with full charge density~\cite{papanikolaou_conceptual_2002}.
For details on the formalism, we refer the reader to Refs.~\cite{lounis_theory_2011,dias_relativistic_2015}

Finally, we note that 
$U_{i}$ in Eq.~\ref{eq:susc-long} denotes the interaction kernel, which in this work is treated in the 
adiabatic local spin-density approximation~\cite{vosko_accurate_1980}.
In the longitudinal case, the kernel $U_{\parallel}$
contains two contributions, namely the Coulomb term
as well as the exchange-correlation (XC) term 
responsible for the many-body effects~\cite{ortenzi_accounting_2012,PhysRevB.96.144410}.
In the transversal case, the Coulomb term drops out leaving only 
the XC term. Importantly, the transverse kernel $U_{\pm} = U_{\mp}$
can be connected to the magnetic moment $m_{0}$ 
and the XC magnetic field $B_{\mathrm{xc}}$ via sum-rules
that allow an accurate description of the dynamical regime: 
for details, we refer the reader to Refs.
~\cite{lounis_dynamical_2010,lounis_theory_2011,dias_relativistic_2015}.

\section{Preliminaries: ground state properties of 3\textit{d} and 4\textit{d} magnetic TM single adatoms deposited on metallic substrates}
\label{sec:gs}

\begin{figure}[t]
\centering
\includegraphics[width=0.95\textwidth]{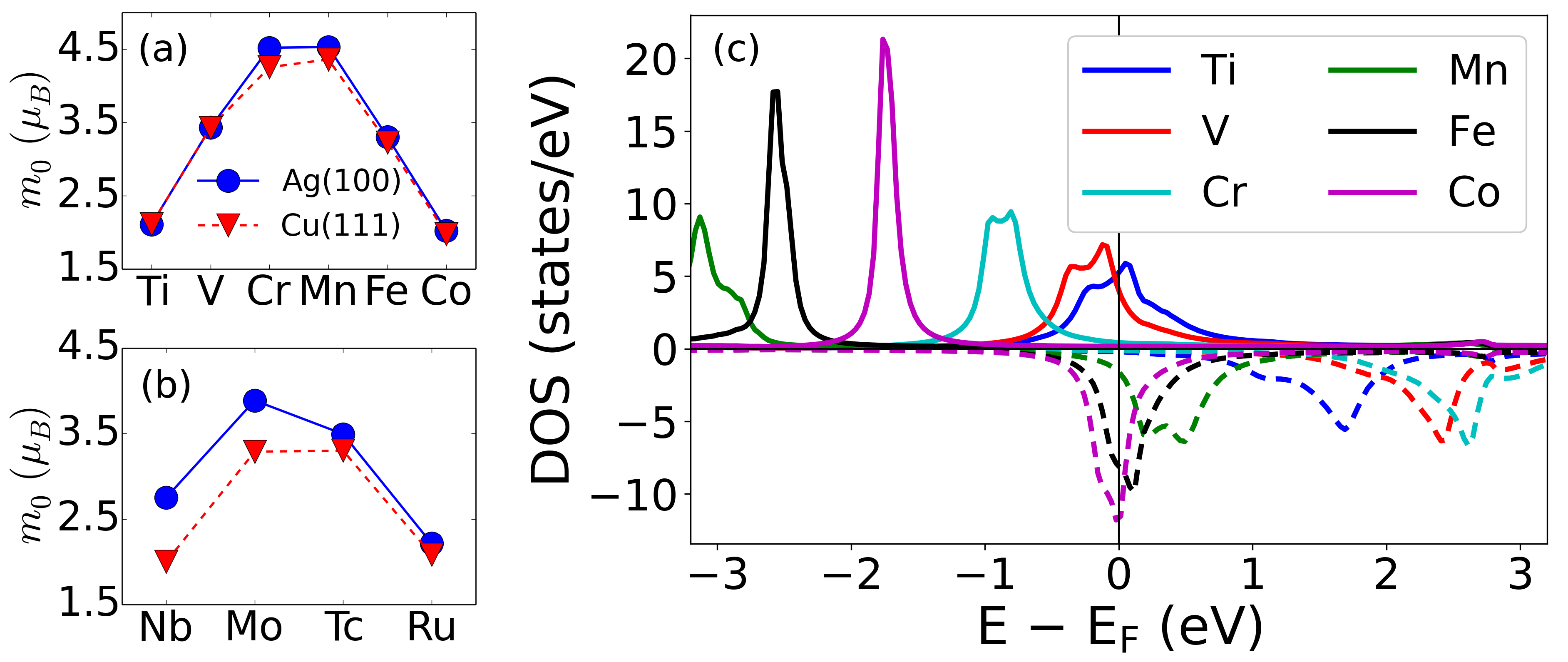}
\caption{Calculated ground state electronic properties of single adatoms. 
(a) Spin magnetic moments for several 3\textit{d} 
transition metal adatoms deposited on Ag(100)  and Cu(111) denoted by circles (blue) and
triangles (red), respectively. 
(b) Same as (a) but for 4\textit{d} adatoms.
(c) Atom-projected total electronic DOS for 3\textit{d} adatoms deposited on Ag(100).
The majority and minority spin channels are denoted by solid (positive) and
dashed (negative) lines, respectively. The vertical line denotes the Fermi level. 
}
\label{fig:M0}
\end{figure}

The main focus of this review is a set of 3\textit{d} and 4\textit{d} adatoms deposited on 
the metallic Ag(100) and Cu(111) substrates.
Before tackling their dynamical properties, we first 
analyze two ground state quantities, namely the spin 
magnetic moment, denoted as $m_{0}$, and the electronic 
DOS.
Let us begin with Fig.~\ref{fig:M0}(a) and \ref{fig:M0}(b), where the calculated  $m_{0}$ is depicted. 
This figure shows that all the considered  adatoms develop large magnetic moments 
of more than $2\,\mu_{\mathrm{B}}$. Furthermore,  $m_{0}$ acquires non-integer values, 
indicating the itinerant character of the 
adatom's \textit{d}-electrons
induced by the hybridization with the electrons of the metallic substrate. 
This feature 
is confirmed by the DOS, which is  displayed in Fig.~\ref{fig:M0}(c) for the specific case of 3\textit{d} 
adatoms deposited on Ag(100).
This figure shows that the \textit{d}-state peaks, so-called virtual bound states, 
are substantially broadened 
(between $\sim0.1$ eV and $\sim1$ eV depending on the adatom),
a well-known consequence of hybridization with 
the substrate~\cite{wildberger_magnetic_1995,Khajetoorians2011,dias_relativistic_2015}.
Another property evidenced  by  Figs.~\ref{fig:M0}(a,b) is that the first of the
atomic Hund's rules is closely fulfilled, i.e.
the half filled \textit{d}-shell elements develop largest magnetic moments, case of
Cr and Mn for 3\textit{d}, Mo and Tc for 4\textit{d}.
Finally, Figs.~\ref{fig:M0}(a,b) show that the choice of metallic substrate and surface orientation
does not substantially affect the spin magnetic moment of the adatom, impliying that 
the substrate's symmetry plays only a minor role in this context. 
We note that these ground state properties are consistent with the original works by 
Dederichs and co-workers~\cite{oswald_giant_1986,wildberger_magnetic_1995,lang_local_1994},
as well as with more 
recent studies~\cite{lounis_dynamical_2010,lounis_theory_2011,dias_relativistic_2015,PhysRevB.91.104420,Schweflinghaus2014,ibanez-azpiroz_zero-point_2016}.

\section{Zero-point spin-fluctuations}
\label{sec:zpsf}

In this section we analyze the ZPSF (Eq.~\ref{eq:fd-theorem}) 
associated to the magnetic single adatoms
presented in Sec.~\ref{sec:gs}.
Two fundamentally different contributions can be distinguished:
the longitudinal $(\xi^{2}_{\parallel})$ and 
transverse $(\xi^{2}_{\perp})$ components of the ZPSF.
In Table \ref{table:zpsf} we have listed the calculated magnitude of both quantities.
Importantly, these values show  
that the longitudinal contribution is approximately an order of magnitude smaller
than the transverse one: the longitudinal component  reaches 
a maximum of $\sqrt{\xi^{2}_{\parallel}}=0.38$ $\mu_{\mathrm{B}}$
for Mo on Ag(100), while $\sqrt{\xi^{2}_{\perp}}>1$ $\mu_{\mathrm{B}}$ for virtually all the adatoms.
This feature reflects the fact that it is much easier to alter the direction of the
magnetic moment than its size. Thus, given that the major contribution to the fluctuations comes
from  the transverse component, 
in the following we will focus on its analysis and its effects on the magnetic stability.

\begin{table}[t]\label{table:zpsf}
\centering
 \begin{tabular}{  c | c | c | c | c | c | c || c | c | c | c | c }
  \hline                       
              & Ti   &  V   &  Cr  &  Mn  &  Fe  &  Co  &  Zr  &  Nb  &  Mo  &  Tc  & Ru  \\ 
  \hline  
  $\sqrt{\xi^{2}_{\parallel}}$ [Ag(100)]  & 0.09 & 0.11 & 0.06 & 0.07 & 0.14 & 0.18 & 0.21 & 0.16 & 0.22 & 0.16 & 0.34 \\
  \hline 
  $\sqrt{\xi^{2}_{\perp}}$ [Ag(100)]  & 2.05 & 2.50 & 3.21 & 3.25 & 3.16 & 1.63 &  0.92 & 2.41 & 2.84 & 3.51 & 1.88 \\
\hline  
\hline 
$\sqrt{\xi^{2}_{\parallel}}$ [Cu(111)]    & 0.12 & 0.14 & 0.12 & 0.10 & 0.08 & 0.14 &  -   & 0.36 & 0.36 & 0.38 & 0.24 \\
         \hline  
$\sqrt{\xi^{2}_{\perp}}$ [Cu(111)]    & 2.49 & 2.80 & 2.31 & 2.56 & 3.02 & 2.57 &  -   & 2.31 & 2.56 & 2.70 & 2.27 \\
         \hline  
    \end{tabular}
     \caption{Calculated longitudinal and transverse contribution to the 
     ZPSF (units of $\mu_{\mathrm{B}}$).}
        \end{table}

\subsection{Connection to spin-excitation spectra}   

\begin{figure}[t]
\centering
\includegraphics[width=0.98\textwidth]{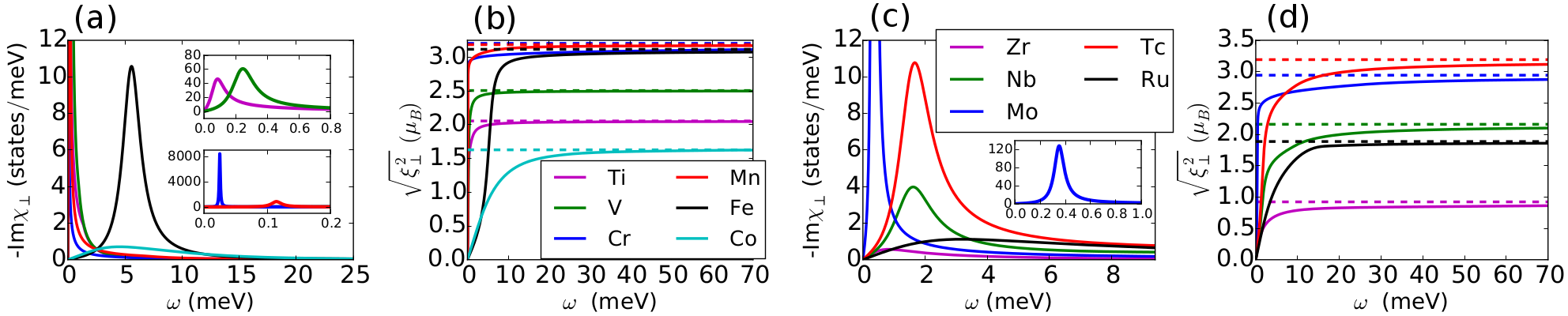}
\caption{Calculated spin-excitation spectra and ZPSF. 
(a) and (c) Density of transverse spin-excitations as given by
$\Im \chi_{\perp}(\omega)$ for selected 3\textit{d} and 3\textit{d} adatoms on Ag(100), respectively. 
Insets address different frequency regions where various resonance frequencies are located.
(b) and (d) Calculated magnitude of the 
mean value of transverse ZPSF for the adatoms
considered in (a) and (c), respectively.
Solid lines depict the evolution of the value
as a function of the upper boundary of the integral of Eq.~\ref{eq:fd-theorem}, while the
horizontal dashed lines represent the converged value.
}
\label{fig:imchi-xi}
\end{figure}

The integrand of Eq.~\ref{eq:fd-theorem} defining the ZPSF is composed by the
imaginary part of the enhanced spin-susceptibility, a 
quantity that displays so-called spin-excitations  associated to the 
damped precessional motion of the spin-moment~\cite{lounis_dynamical_2010,lounis_theory_2011,dias_relativistic_2015}.
Calculated transverse spin-excitation and spin-fluctuation spectra
for TM adatoms deposited on Ag(100) are shown in Fig.~\ref{fig:imchi-xi}. 
Let us first   focus on Figs.~\ref{fig:imchi-xi}(a,c), which 
display $\Im\chi_{\perp}(\omega)$ for 3\textit{d} and 4\textit{d} adatoms, respectively. 
These figures reveal
the existence of a large peak in the meV range for all adatoms, 
corresponding to a spin-excitation.
Note that the  associated resonance frequency 
is dependent on the MAE, which is in turn settled by the SOC~\cite{dias_relativistic_2015}.
The largest resonance frequency of $\sim6$ meV pertains to the Fe adatom, 
while the rest of adatoms display the peak at $\lesssim 4$ meV; in the particular cases of
Ti, V, Mn, Cr and Mo, we find the spin-excitation at extremely small frequencies, $\lesssim 0.5$ meV,
impliying that the net effect of SOC is very weak in these adatoms. 
Figs.~\ref{fig:imchi-xi}(a,c) display yet another feature, namely the width of the spin-excitation peak, 
which is linked to the amount of electron-hole
Stoner excitations near the Fermi level~\cite{PhysRevB.91.104420}:
Ru and, to some extent also Nb and Tc, 
display large widths as compared to the rest of adatoms, in particular
those that peak below 1 meV.

We next  turn to the magnitude of the transverse ZPSF (Eq.~\ref{eq:fd-theorem}),
as displayed in Figs.~\ref{fig:imchi-xi}(b,d)  
for 3\textit{d} and 4\textit{d} adatoms on Ag(100), respectively; 
the solid lines depict the evolution 
of the mean value, $\sqrt{\xi_{\perp}^{2}}$, as a function of the upper 
boundary of the frequency integral,
whereas the converged value
is marked by the horizontal dashed lines. 
As already pointed in the discussion of  Table \ref{table:zpsf},   
$\sqrt{\xi_{\perp}^{2}}$ is of the order of the Bohr magneton;
the elements with largest values are Cr, Mn, Fe and Tc, 
with $\sqrt{\xi_{\perp}^{2}}\sim3\,\mu_{\mathrm{B}}$,
whereas in Zr and Co this value is reduced by more than 50\%.
As the calculations demonstrate, the main contribution to the integral of 
Eq.~\ref{eq:fd-theorem} comes from the spin-excitation peak in the meV region,
which represents between 70\% and $\sim$100\% of the total depending on the adatom.

\begin{figure}[t]
\centering
\includegraphics[width=0.75\textwidth]{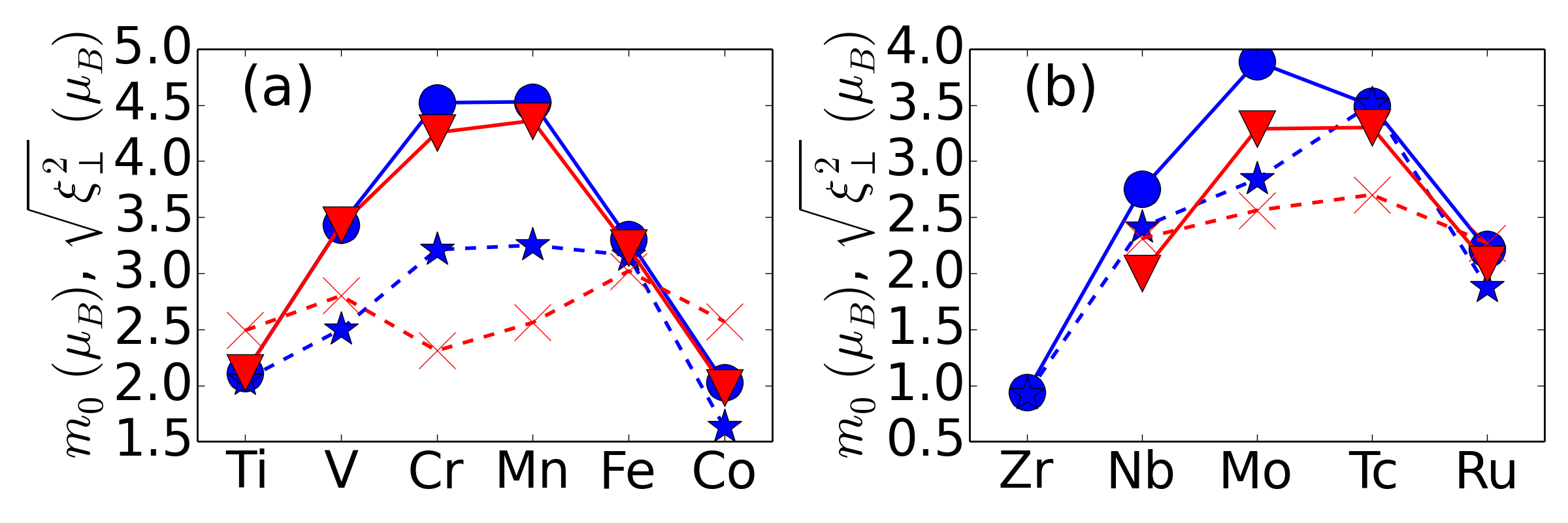}
\caption{ Comparison of local magnetic moment and mean value of the transverse 
ZPSF. 
(a) 3\textit{d} adatoms and  (b) 4\textit{d} adatoms. As in Figs. \ref{fig:M0}(a,b),
blue circles and red triangles denote $m_{0}$ for adatoms deposited on Ag(100)
and Cu(111), respectively. 
In addition, blue stars and red crosses denote $\sqrt{\xi^{2}_{\perp}}$
for adatoms deposited on Ag(100) and Cu(111), respectively.}
\label{fig:M0-xi}
\end{figure}

Figs.~\ref{fig:M0-xi}(a) and \ref{fig:M0-xi}(b) show an instructive comparison between 
the  magnitude of the transverse ZPSF and the
local magnetic moments of the adatoms. 
These figures demonstrate that $\sqrt{\xi^{2}_{\perp}}$ (blue stars) 
is always an appreciable fraction of $m_{0}$ (blue circles); 
one finds $\sqrt{\xi^{2}_{\perp}}\sim m_{0}/2$ for Mn and Cr, while  
$\sqrt{\xi^{2}_{\perp}}\sim m_{0}$ for Ti, Fe, Zr, Nb and Ru,  
revealing that the latter suffer from strong deviations
of the direction of the local magnetic moment.
We also note that both $m_{0}$ and $\sqrt{\xi^{2}_{\perp}}$ 
follow the evolution predicted by Hund's rules, i.e.~adatoms with nearly half-filled \textit{d}-shells have largest values: Cr and Mn among 3\textit{d}, 
and Mo and Tc among 4\textit{d}.
This trend is clearly fulfilled in the case of the magnetic moment as already mentioned in Sec. \ref{sec:gs},
while the evolution of the fluctuations shows some exceptions like Fe and Tc.
Noteworthily, a similar relationship between magnetic moment and
fluctuations has been
very recently found also for bulk transition metals~\cite{PhysRevB.96.184418}.
In addition, Fig.~\ref{fig:M0-xi} also reveals that the 
spin-fluctuation-to-magnetization ratio (SFMR) is generally larger
in 3\textit{d} adatoms than in 4\textit{d}.
As a final remark, let us note that Mo, which is 
the only 4\textit{d} adatom deposited on Ag(100) 
that shows an experimentally measurable magnetic signal~\cite{mo-ag100},
has by far the lowest 
SFMR among 4\textit{d} elements.

Figs.~\ref{fig:M0-xi}(a) and \ref{fig:M0-xi}(b) also display an extension of the analysis
to the  same set of adatoms deposited on Cu(111),
which exhibit essentially the same features 
as on Ag(100). This again
indicates that symmetry plays a minor role in these largely hybridized 
metallic systems with strong itinerant character. 
The only noteworthy difference is that the SFMR of Cr and Mn 
is somewhat lower than what is expected from the trend of Fig.~\ref{fig:M0-xi};
we will discuss the origin of this feature shortly.
Let us also note that Co adatoms 
show a very large SFMR in both Ag(100) and Cu(111) (see Fig.~\ref{fig:M0-xi}).
Incidentally, Co is known to behave as
a Kondo system when deposited on various metallic substrates~\cite{madhavan_tunneling_1998,manoharan_quantum_2000,
PhysRevLett.94.036805}.
Since both ZPSF and the Kondo effect are driven by spin-flip excitations with substrate electrons
(see discussion below and Fig.~\ref{fig:LLG-xi-eta-wres}), 
it is tempting to associate both effects. However, only a more advanced theory capable of 
describing such many-body effects can prove this hypothesis.

\subsection{Interpretation in terms of the Landau-Lifshitz-Gilbert model}

In the following, we turn to  identify the fundamental factors that determine the 
magnitude of the transverse ZPSF.
For this purpose, we consider  the 
Landau-Lifshitz-Gilbert  (LLG)  equation~\cite{gilbert_phenomenological_2004}.
The LLG is widely employed for characterizing the spin-dynamics of macroscopic
magnetic systems, and more recently also for microscopic systems~\cite{PhysRevB.91.104420,dias_relativistic_2015}
as it allows to accurately reproduce the \textit{ab initio} calculations 
by extracting the relevant parameters 
(see also Ref.~\cite{PhysRevLett.110.147201} for a derivation of 
the quantum counterpart of the LLG equation). 
The details of the model are worked out in Appendix \ref{appendix:LLG}.
Here, let us recall that  within the LLG model, 
the imaginary part of the transverse spin-susceptibility takes the form
of an skewed Lorentzian, i.e.
\bek\label{eq:LLG-susc}
\text{Im}\chi_{\pm}^{\text{LLG}}(\omega)=\dfrac{M\gamma}{2(1+\eta^{2})}
\cdot\dfrac{\eta\omega}{(\omega-\omega_{0})^{2}+(\eta\omega_{0})^{2}}.
\ek 
The parameters entering this model are
the Gilbert damping, $\eta$, which determines the width of the spin-excitation peak and is dominated by Stoner excitations, and the resonance frequency,  
$\omega_{\text{res}}=\omega_{0}\sqrt{1+\eta^{2}}=\gamma B_{\text{eff}}/\sqrt{1+\eta^{2}}$,
where $\gamma$ is the gyromagnetic ratio
and $B_{\text{eff}}$ an effective magnetic field whose magnitude is determined by 
the strength of SOC (and of the external field, if considered); 
a detailed discussion can be found in, e.g., Ref.~\cite{dias_relativistic_2015}.

The integral of Eq.~\ref{eq:fd-theorem} can be calculated analytically within the LLG model and yields
immediately the  expression for the transverse ZPSF:
\begin{equation}
\label{eq:LLG-xi}
\xi^{2}_{\text{LLG}}=
\dfrac{M\gamma}{\pi(1+\eta^{2})}
\Big(
\dfrac{\eta}{2} \log\dfrac{(x^{2}+\eta^{-2}+1)^{2}-(2x\eta^{-1})^{2}}{(\eta^{-2}+1)^{2}} + 
\vartheta(x,\eta)
\Big).
\end{equation}
Above, $\vartheta(x,\eta)=\arctan (x-\eta^{-1}) - 
\arctan (x+\eta^{-1}) + 2\arctan \eta^{-1}$
and $x=\omega_{c}/\eta\omega_{0}$,
with $\omega_{c}$ a 
cutoff frequency to be converged.

\begin{figure}[t]
\centering
\includegraphics[width=0.7\textwidth]{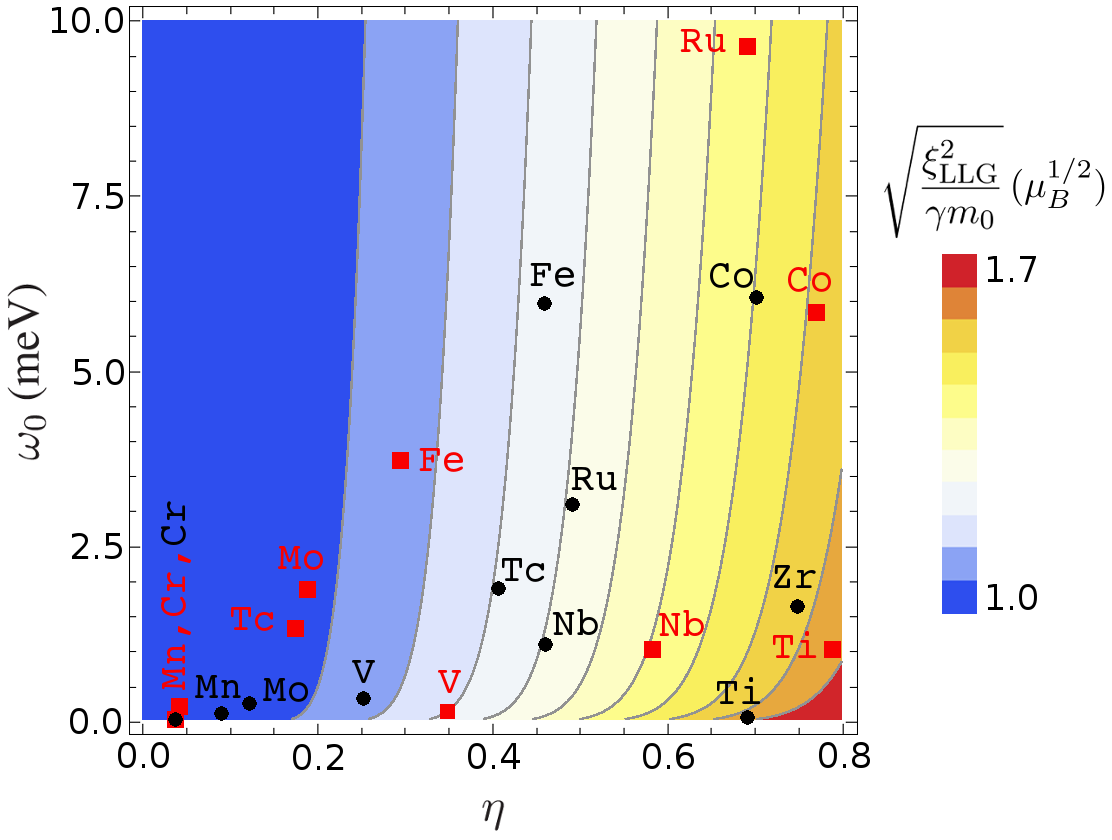}
\caption{The background shows a 2D plot of the 
SFMR as a function of the damping
and the resonance frequency in the LLG model (Eq.~\ref{eq:LLG-xi}). 
Circles (black) and squares (red) denote various adatoms 
deposited on Ag(100) and Cu(111),
respectively, where the parameters $\eta$ and $\omega_{0}$ (see discussion of Eq.~\ref{eq:LLG-susc}) have been extracted from a fit to
the \textit{ab initio} spin-susceptibility.}
\label{fig:LLG-xi-eta-wres}
\end{figure}

Eq.~\ref{eq:LLG-xi} provides an interpretation
for the magnitude of the fluctuations in terms of the physical parameters of the LLG model, which
are in turn related to the electronic structure of the adatoms.
In particular, three major ingredients come into play:
$m_{0}$, $\omega_{0}$ and $\eta$.
First of all, Eq.~\ref{eq:LLG-xi} reveals that $\sqrt{\xi^{2}_{\text{LLG}}}$ is proportional to $\sqrt{m_{0}}$,
which already accounts for 
the rough proportionality displayed by the \textit{ab initio} calculations (see Fig.~\ref{fig:M0-xi}).
Moreover, this relationship further indicates that fluctuations are 
relatively weaker for adatoms with large magnetic moments, namely
the nearly half-filled \textit{d}-shell elements.  
Interestingly, the simple relationship of Eq.~\ref{eq:LLG-xi}  allows to analyze the SFMR,  
$\sqrt{\xi^{2}_{\text{LLG}}/\gamma m_{0}}$, 
as a function of $\eta$ and $\omega_{0}$ by setting the standard value 
for the gyromagnetic ratio $\gamma=2$; the  map obtained by doing so is shown in the background of 
Fig.~\ref{fig:LLG-xi-eta-wres}.
This figure evidences that the SFMR
is very much enhanced by the damping since $\sqrt{\xi^{2}_{\text{LLG}}/\gamma m_{0}}$ varies by almost  
70\% in the range of values considered for $\eta$.
In contrast, Fig.~\ref{fig:LLG-xi-eta-wres} shows that the transverse ZPSF are reduced for large
resonance frequencies, but the induced variation is much less important than that of the damping.
In order to quantitatively connect these features to  \textit{ab initio} calculations, 
we have extracted the parameters $\eta$ and  $\omega_{0}$ by fitting  
$\text{Im}\chi_{\pm}^{\text{LLG}}(\omega)$ to the
calculated spin-susceptibility (e.g., Figs.~\ref{fig:imchi-xi}(a,c))
for all adatoms.
In this way, we locate the position of each adatom on the map
of Fig.~\ref{fig:LLG-xi-eta-wres}, 
as depicted by the circles and squares for the case of the Cu(111) and Ag(100) substrates, respectively.
The resulting distribution evidences that the origin of the 
large SFMR found in the case of 
Ti, Co, Ru and Nb on both substrates, as well as Fe and Zr on Ag(100),
is mainly due to the large damping factors of these adatoms, $\eta\gtrsim0.5$, while
this tendency is only slightly modified by the position of the resonance frequency.

We note that a remarkable feature is revealed when $\eta$ tends to zero, i.e.,
\bek\label{eq:zero-damping}
\lim_{\eta\rightarrow 0}\sqrt{\dfrac{\xi^{2}_{\text{LLG}}}{\gamma m_{0}}}=1 \;\mu_{\mathrm{B}}^{1/2}\Rightarrow
\lim_{\eta\rightarrow 0}\sqrt{\xi^{2}_{\text{LLG}}}=\sqrt{\gamma m_{0}}.
\ek
The expression above shows that the transverse ZPSF remain finite even in the extreme case of a spin-excitation with vanishing width.
This intrinsic contribution is required by Heisenberg's uncertainty principle, since different spin-operators do not commute with each other~\cite{clerk_introduction_2010}, thus setting a lower boundary for the fluctuations.
This is indeed the case found for Mn, Cr and Mo adatoms, which, despite their small resonance frequency, their tiny damping ($\eta\lesssim 0.05$) makes $\sqrt{\xi^{2}_{\text{LLG}}/\gamma m_{0}}$ approach the intrinsic minimum value for these elements.
Interestingly, we note that Eq.~\ref{eq:zero-damping} represents a sensible limit to give a rough estimate for the ZPSF of adatoms deposited on insulating and semiconducting substrates, since the substrate hybridization giving rise to the damping tends to be far smaller on those type of substrates as compared to the metallic ones considered in this review.
Let us illustrate this notion by considering the important example of a Ho atom deposited on ultrathin MgO(100) layers grown on Ag(100), which experimentally exhibits a stable magnetic moment for time-periods larger than $10^{3}$ s~\cite{donati_magnetic_2016}.
From Hund's rules the Ho adatom should have a magnetic moment of $10\,\mu_{\mathrm{B}}$, as found experimentally~\cite{natterer_reading_2017}.
Under this assumption, a quick estimate using Eq.~\ref{eq:zero-damping} reveals that the magnitude of the ZPSF in Ho should be $\sim4.5\,\mu_{\mathrm{B}}$.
This, in turn,  means that the resulting SFMR is nearly a factor of 2 smaller than in the elements with lowest SFMR studied in this review (see Eq.~\ref{eq:mae-xi} and Fig.~\ref{fig:MAE}), implying that the relative impact of the quantum spin-fluctuations for Ho on MgO/Ag(100) is particularly weak.

\subsection{Impact on the magnetic anisotropy energy}

As a last step in the analysis of this section devoted to spin-fluctuations, we turn now to analyze 
the effects of the transverse ZPSF on the magnetic stability of the adatoms.
For this  purpose, we estimate how the fluctuations affect the MAE,
the quantity defining the height of the energy barrier when the magnetic moment is made to rotate away from its easy axis. 
We note that the MAE  is closely connected to the resonance frequency 
of the spin-excitations (see, e.g., Ref.~\cite{dias_relativistic_2015})
that give rise to the primary contribution to the transverse ZPSF; 
most importantly, the energy scale of both quantities is the same (meV).

\begin{figure}[t]
\centering
\includegraphics[width=0.75\textwidth]{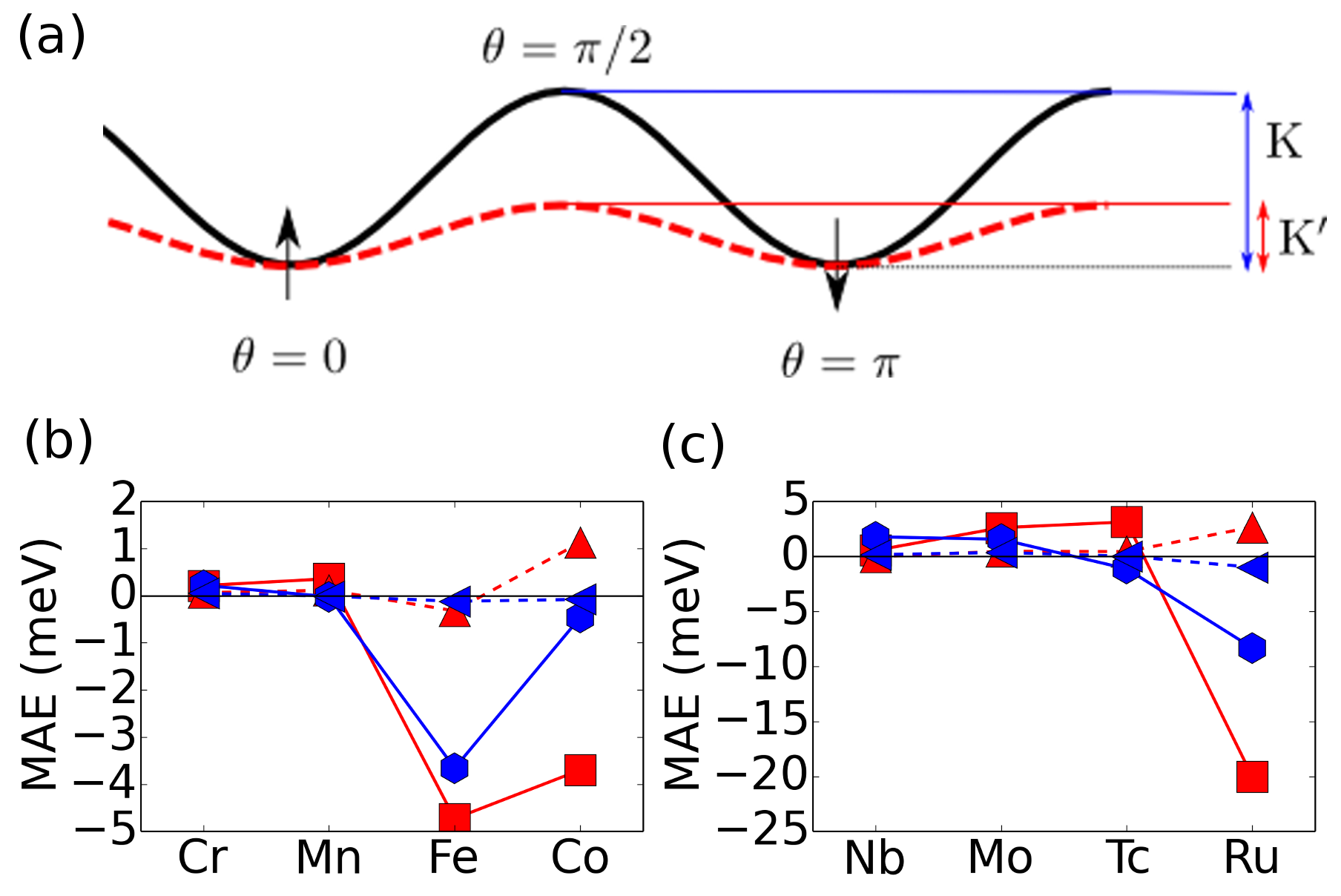}
\caption{(a) Schematic illustration of the renormalization of the MAE induced by the
transverse ZPSF. 
K and K$'$ respectively represent the static and renormalized anisotropy constant
(see Eq.~\ref{eq:mae-xi}). (b) and (c) Calculated anisotropy constants for 3\textit{d}
and 4\textit{d} adatoms, respectively. Blue hexagons and leftward triangles denote 
K and K$'$  for adatoms on Ag(100), respectively, while red squares and upward triangles
denote K and K$'$  for adatoms on Cu(111), respectively.
}
\label{fig:MAE}
\end{figure}

We consider the following expression for the MAE for uniaxial systems, 
\bek
E_{a}(\theta)=\text{K}\dfrac{(\textbf{m}\cdot\hat{\textbf{e}}_{z})^{2}}{\textbf{m}^{2}}=
\text{K}\cos^{2}\theta,
\ek
where K is the so-called anisotropy constant.
As a consequence, the energy barrier between the magnetic moment pointing 
along the $z$ axis ($\theta=0$) and a perpendicular axis ($\theta=\pi/2$)
is given by
$E_{a}(\theta=0)-E_{a}(\theta=\pi/2)=\text{K}$,
as it is schematically illustrated in Fig. \ref{fig:MAE}(a).
In the spirit of the spin-fluctuation theory of Moriya~\cite{moriya-book}, 
we now let magnetic moment fluctuate around its equilibrium value,
i.e. $\textbf{m}^{2}\rightarrow 
 \left( 
m_{0}\hat{\textbf{e}}_{z}+\sum_{\perp}\boldsymbol{\xi}_{\perp}
\right)^{2} $.
Introducing this term into the definition of $E_{a}(\theta)$,
we obtain a renormalized expression for the MAE, i.e.
\bek
E_{a}(\theta,\xi^{2}_{\perp})=\dfrac{\text{K}(m_{0}^{2}\cos^{2}\theta+\xi^{2}_{\perp}\sin^{2}\theta)}{m_{0}^{2}+2\xi^{2}_{\perp}}.
\ek
In practice, this implies that the energy barrier gets effectively reduced by the transverse ZPSF,
\begin{equation}
\label{eq:mae-xi}
E_{a}(\theta=0,\xi^{2}_{\perp})-E_{a}(\theta=\pi/2,\xi^{2}_{\perp}) = 
\text{K}\left(
1-\dfrac{3\xi^{2}_{\perp}}{m_{0}^{2}+2\xi^{2}_{\perp}}\right)\equiv \text{K}'.
\end{equation}
The above is characterized by a modified anisotropy 
constant K$'$, also illustrated in Fig.~\ref{fig:MAE}(a).

In Figs.~\ref{fig:MAE}(b) and \ref{fig:MAE}(c) we show the calculated values for  
the anisotropy constant, K, as well as for the reduced one, K$'$: the former has been evaluated 
by band energy differences following the magnetic 
force theorem~\cite{PhysRevB.32.2115}, while the latter has been calculated using the
values of the magnetic moments and transverse ZPSF of the adatoms.
These figures show a large and generalized 
reduction of the MAE, which goes
beyond  estimates that do not take  into account fluctuation effects~\cite{dias_relativistic_2015}. 
In cases where the ZPSF are larger than $m_{0}$ (mostly 4\textit{d} adatoms),
the magnetic moment direction becomes destabilized, 
as reflected by the sign change in the renormalized MAE. 
Let us pay special attention to the well-studied  case of Fe on Cu(111), where 
the MAE is reduced  from $\mathrm{K} =-4.73$ meV to K$'=-0.32$ meV. Interestingly, this means that 
ZPSF considerably remedy the disagreement with the 
experimental MAE, which is nearly $\sim-1$ meV~\cite{Khajetoorians2011,PhysRevB.91.235426}.
More importantly, this example shows that the renormalization of the MAE
due to the ZPSF has the correct order of magnitude.
Focusing next on the 4\textit{d} elements Ru and Nb on Ag(100), 
we find that the reduction of the MAE predicted by the calculations 
is notably large, i.e. $\gtrsim90\%$.
It is noteworthy that these two cases have been thoroughly  studied experimentally, 
which have concluded that there is no sizable magnetic signal in the systems~\cite{honolka_absence_2007,schafer-nb}. 
This suggests that the ZPSF might also play a major role
in the experimentally observed absence of stable magnetism.

\section{Spin-relaxation times}
\label{subsec:relax}

While the analysis of Sec.~\ref{sec:zpsf} deals with the magnitude
of the fluctuations, this section instead focuses on the time scales.
For this purpose, 
we consider two particular relaxation processes of magnetic
single adatoms, namely the longitudinal and transverse 
spin-relaxations characterized by the relaxation times $T_{\parallel}$ and $T_{\perp}$, respectively. 
Physically, 
$T_{\parallel}$ characterizes the relaxation of the size of the adatom's spin magnetic moment
while $T_{\perp}$ describes its damped precessional motion.
In the following, we analyze each of this processes separately.

\subsection{Longitudinal component}
\label{subsec:long}

In essence, the longitudinal spin-susceptibility 
$\chi_{\parallel}(\omega)$ (see Eq.~\ref{eq:susc-long}) 
describes the ability of the system
to continuously modify the size of its magnetic moment by an externally applied
time-dependent magnetic perturbation along the magnetization direction.
The dynamics of this process can be 
phenomenologically studied in terms of the longitudinal Bloch equation, 
which yields the following form for the enhanced spin-susceptibility~\cite{white_quantum_2007}
(see Appendix \ref{appendix:Bloch-long}),
\bek\label{eq:bl-Tlong}
\chi^{\text{Bl}}(\omega) =
\dfrac{\chi^{\text{Bl}}_{0}}{1-\mathrm{i}\,\omega\,T_{\parallel}}.
\ek
Above, $\chi^{\text{Bl}}_{0}$ denotes a static spin-susceptibility, while 
$T_{\parallel}$ corresponds to the longitudinal relaxation time mentioned in the introduction above. 
The aim is to establish a direct comparison between 
Eqs.~\ref{eq:susc-long} and \ref{eq:bl-Tlong}. For this purpose, let us
use the first-order Taylor expansion of the KS spin-susceptibility~\cite{PhysRevLett.119.017203} 
\bek\label{eq:chi-KS-long}
\chi^{\mathrm{KS}}_{\parallel}(\omega)\simeq \rho_{\mathrm{F}}   
-\mathrm{i}\,n_{\mathrm{e-h}}\,\omega,
\ek
with $\rho_{\mathrm{F}}=\rho_{\mathrm{F},\uparrow}+\rho_{\mathrm{F},\downarrow}$ the DOS at the Fermi level
and $n_{\mathrm{e-h}}=\pi(\rho^{2}_{\mathrm{F},\uparrow}+\rho^{2}_{\mathrm{F},\downarrow})/2$ the density of electron-hole
excitations of the same spin channel.
By inserting $\chi^{\mathrm{KS}}_{\parallel}(\omega)$ of Eq.~\ref{eq:chi-KS-long}
into Eq.~\ref{eq:susc-long},
$\chi_{\parallel}(\omega)$ acquires a functional form in $\omega$
equal to that of $\chi^{\text{Bl}}(\omega)$ in Eq.~\ref{eq:bl-Tlong}. This then allows to obtain an 
expression for the longitudinal relaxation time in terms of
basic electronic properties  
(see Appendix \ref{appendix:Bloch-long} for details):
\begin{equation}
\label{eq:Tlong}
T_{\parallel}=
\dfrac{U_{\parallel}\,n_{\mathrm{e-h}}}{U_{\parallel}\,\rho_{\mathrm{F}}-1}.
\end{equation}
The above expression  shows that the longitudinal relaxation time is 
mostly settled by the magnitude of electron-hole excitations weighted by
the XC kernel (see the numerator of Eq.~\ref{eq:Tlong}); this product is of
order 1/eV, hence settling the overall time scale of 
$T_{\parallel}$ as fs.
Furthermore, Eq.~\ref{eq:Tlong} shows that $T_{\parallel}$ diverges as 
$U_{\parallel}\,\rho_{\mathrm{F}}\rightarrow 1$ 
(see the unitless denominator
in the equation) 
and hence its magnitude can be strongly modified in this limit,
i.e.~as the system approaches the magnetic transition point.
In the following,
we first focus on quantitatively 
analyzing the ingredients of Eq.~\ref{eq:Tlong}
and subsequently turn to $T_{\parallel}$ itself.

\begin{figure}[t]
\centering
\includegraphics[width=0.7\textwidth]{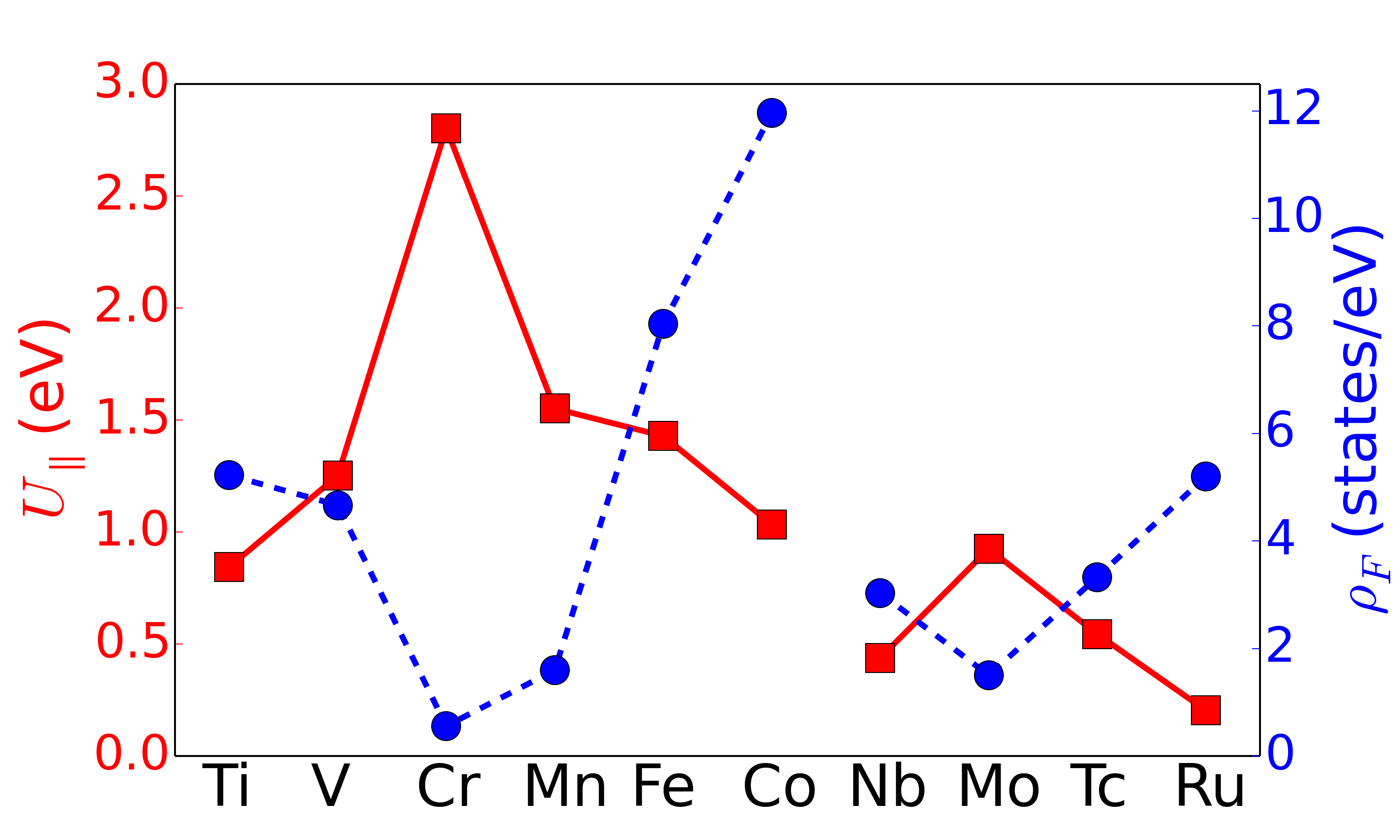}
\caption{Calculated longitudinal kernel and DOS at $E_{F}$  for 3\textit{d} and 4\textit{d} magnetic adatoms deposited on Ag(100). 
Solid (red) and dashed (blue) lines show the 
values for $U_{\parallel}$ (Eq.~\ref{eq:U_long}) and 
$\rho_{\mathrm{F}}$
whose corresponding ordinate axes are placed on the left (red) and right (blue)
of the graph, respectively. Note that lines are broken in order to separate
3\textit{d} from 4\textit{d} elements.
}
\label{fig:Udotrho}
\end{figure}

In order to compute reliable values for the
kernel $U_{\parallel}$, one can  make use of 
the static limit of Eq.~\ref{eq:susc-long}, from which
\bek\label{eq:U_long}
U_{\parallel} = \rho_{\mathrm{F}}^{-1}-\chi^{-1}_{\parallel}(0).
\ek
We note that $\chi_{\parallel}(0)$ can be calculated by a standard ground state DFT calculation
with a static magnetic field $\Delta B$ via $\chi_{\parallel}(0)=\Delta m /\Delta B$,
with $\Delta m$ the corresponding self-consistent change of the magnetic moment~\cite{kubler_theory_2009}. 
In Fig.~\ref{fig:Udotrho} we show the calculated values of $\rho_{\mathrm{F}}$ and $U_{\parallel}$
for several 3\textit{d} and 4\textit{d} adatoms deposited on
Ag(100). The most important message conveyed by this figure is the large variation
of $U_{\parallel}$ among different elements; while  $U_{\parallel}\lesssim 0.5$ eV
for most 4\textit{d} elements, $U_{\parallel}\gtrsim 1.5$ eV for various 3\textit{d} elements,
reaching a maximum of one order of magnitude difference between Ru and Cr. 
A second important feature revealed by Fig.~\ref{fig:Udotrho} 
is the distribution of $U_{\parallel}$ within each \textit{d}-shell, whereby
it is smallest at the ends of the row  
--- case of Ti and Co among 3\textit{d}, Nb and Ru among 4\textit{d} --- 
and highest in the middle of the row 
--- case of Cr and Mn among 3\textit{d}, Mo and Tc among 4\textit{d} ---,
yielding an approximate inverted V-shape.
We note that $\rho_{\mathrm{F}}$ in Fig.~\ref{fig:Udotrho} shows the opposite behavior, 
i.e.~it is minimum for Cr and maximum for Co and Ru.
This is consistent with Eq.~\ref{eq:U_long}, although we note strong deviations
from the $U_{\parallel}\propto \rho_{\mathrm{F}}^{-1}$ relationship 
(see in particular the case of Cr), 
revealing the importance
of the term $\chi^{-1}_{\parallel}(0)$ in Eq.~\ref{eq:U_long}.

\begin{figure}[t]
\centering
\includegraphics[width=0.95\textwidth]{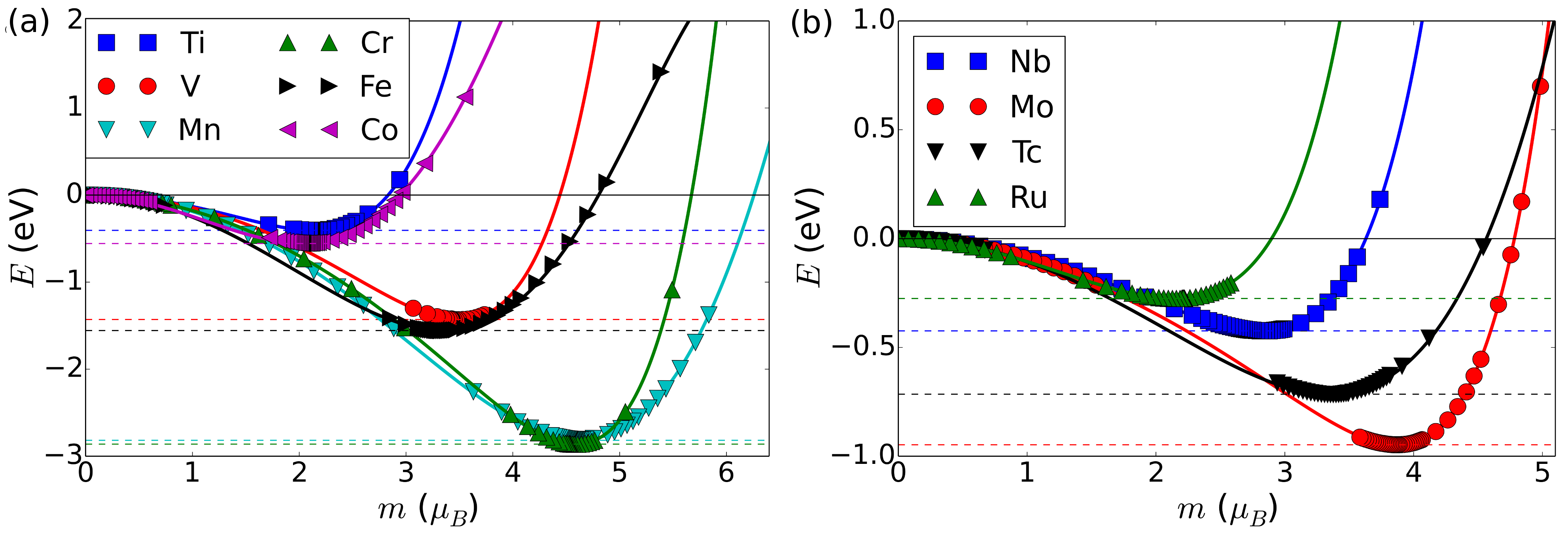}
\caption{Calculated energy as a function of the adatom's magnetic moment. (a) and (b)  
3\textit{d} and 4\textit{d} elements, respectively. DFT calculations are denoted by markers,
while the solid lines are a fit to the calculations using 
$E=\sum_{i=1}^{4} a_{i}m^{2i}$.
The horizontal dashed lines 
mark the minimum energies for different adatoms.
}
\label{fig:EM}
\end{figure}

The inverse of the static spin-susceptibility  
$\chi^{-1}_{\parallel}(0)$ is closely connected to the magnetic equation of state,
i.e.~the dependence of the energy $E$ as a function of the magnetic moment $m$,
via~\cite{kubler_theory_2009},
\bek\label{eq:E(m)}
\chi^{-1}_{\parallel}(0)= \dfrac{\partial^{2}E(m)}{\partial m^{2}}\Big|_{m=m_{0}}<0.
\ek
In essence, the magnetic equation of state informs about how
stable the magnetic solution is in comparison to the non-magnetic one.
Fig.~\ref{fig:EM} shows the calculated $E(m)$ for the set of adatoms 
considered in Fig.~\ref{fig:Udotrho} (calculations were carried out 
by employing DFT for a series of external magnetic fields~\cite{kubler_theory_2009}). This figure
reveals that the energy difference between the non-magnetic and magnetic state,
\bek
\Delta E = E(m=0) - E(m=m_{0}),
\ek
is of the order of eV and can largely vary for different adatoms.
Importantly, the calculations show that 3\textit{d} adatoms overall have a substantially 
larger $\Delta E$ than 4\textit{d} adatoms.
For Cr, for instance, $\Delta E\sim 3$ eV, while for Ru $\Delta E\sim 0.25$ eV.
Therefore, DFT predicts most 3\textit{d} adatoms to be magnetically more stable than 4\textit{d} ones,
as expected.
Furthermore, given that Eq.~\ref{eq:E(m)} together with Eq.~\ref{eq:U_long} 
relates the XC kernel to the second derivative of the
equation of state at $M_{0}$, one can establish an approximate connection between
the depth of the minimum of $E(M)$ and the value of $U_{\parallel}$,
as it is visible from the comparison of Figs.~\ref{fig:EM} and \ref{fig:Udotrho};
the deeper the minimum, the larger $U_{\parallel}$.

\begin{figure}[t]
\centering
\includegraphics[width=0.95\textwidth]{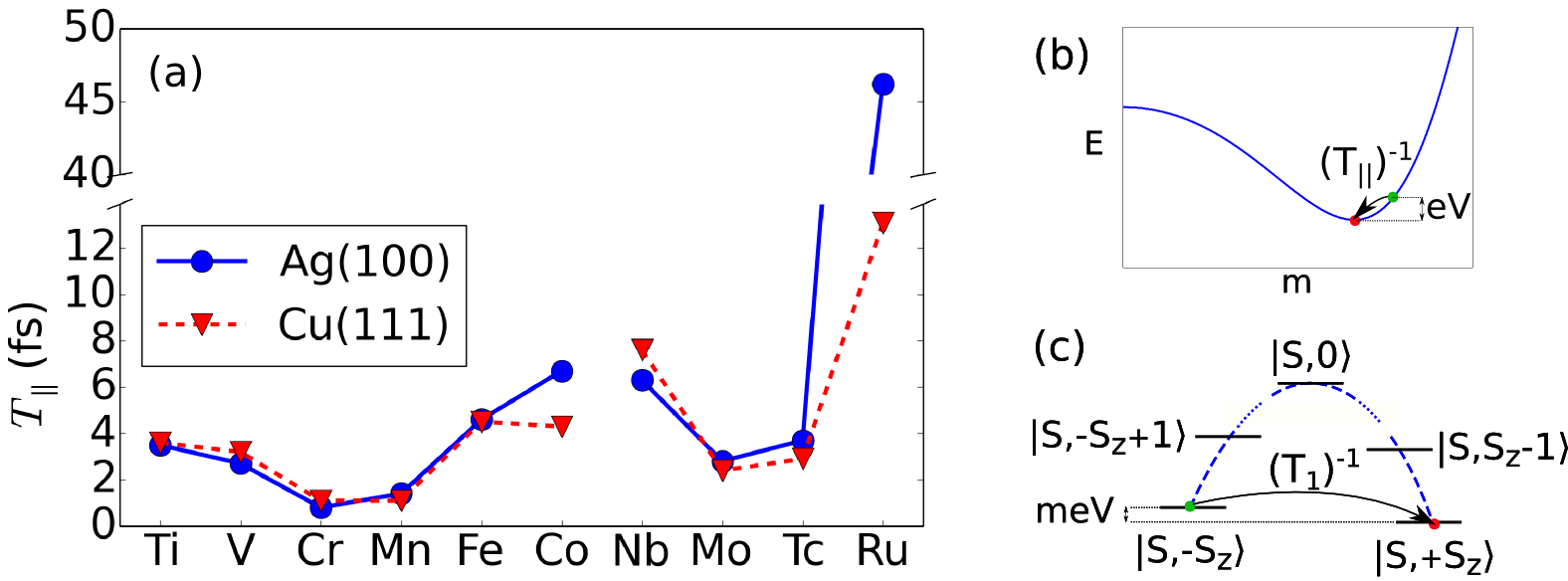}
\caption{Longitudinal relaxation time. (a) Calculated values for the longitudinal relaxation time 
$T_{\parallel}$ for magnetic 3\textit{d} and 4\textit{d} 
    adatoms on Ag(100) (blue circles) and Cu(111) (red triangles).
(b) and (c) Schematic illustration
of two different processes contributing to 
the spin-relaxation time.
Green and red dots respectively represent the excited and ground states.
(b) Schematic description of the 
excitations contributing to $T_{\parallel}$
considered in this work. These take place in a continuum energy landscape of 
order eV and are driven by direct spin-conserving electron-hole transitions.
(c) Illustration of the prototypical energy diagram  
used to describe the experiments of 
Refs.~\cite{loth_measurement_2010,rau_reaching_2014,baumann_electron_2015,paul_control_2017}.
The initially degenerate two ground states of maximum $S_{z}$ become non-degenerate
by energy $\sim$ meV under externally applied magnetic fields of 1--10 T.
Transitions between these two states then determine $T_{1}$,
which take place via quantum tunneling. 
}
\label{fig:Tparallel}
\end{figure}

We note that following the above procedure, one can also extract the 
kernel in the non-magnetic ground state,  
i.e.~the so-called Stoner XC parameter $I_{\mathrm{xc}}$~\cite{janak_uniform_1977}.
This can be achieved by considering the curvature of $E(M)$ not at $m=m_{0}$ but at
$m={0}$, as well as using the non-magnetic DOS in Eq.~\ref{eq:U_long}
instead of the magnetic one. As it is clearly visible from Fig.~\ref{fig:EM}, 
the curvature is very different at $m={0}$ and $m=m_{0}$. 
Furthermore, $\rho_{\mathrm{F}}$ can also strongly vary from a magnetic to a non-magnetic calculation.
As a consequence, the distribution of $I_{\mathrm{xc}}$ along the transition metal series
first reported by Janak in Ref.~\cite{janak_uniform_1977}
is very different
to that of $U_{\parallel}$ illustrated in Fig.~\ref{fig:Udotrho}.

Having analyzed the properties of
$\rho_{\mathrm{F}}$ and $U_{\parallel}$, we next focus on the longitudinal relaxation time $T_{\parallel}$.
The values calculated from Eq.~\ref{eq:Tlong}
are plotted in Fig.~\ref{fig:Tparallel}(a) for 3\textit{d} and 4\textit{d} adatoms deposited on Ag(100) and Cu(111). 
$T_{\parallel}$ is of the order of a few fs in all cases, being overall
slightly larger for 4\textit{d} than 3\textit{d} adatoms, while the choice of substrate
does not substantially affect it. 
Within each \textit{d}-shell, $T_{\parallel}$ is largest at the ends of the row 
while it is minimum for the half filled elements, thus resembling
the behavior of $\rho_{\mathrm{F}}$ (compare Figs.~\ref{fig:Udotrho} and \ref{fig:Tparallel}(a)).
Ru on Ag(100) has the highest value of 
$T_{\parallel}\sim 50$ fs, mainly as a consequence of the denominator of 
Eq.~\ref{eq:Tlong} being closer to zero than in other elements. 
In contrast,
Cr and Mn have $T_{\parallel}\sim 1$ fs in both substrates, i.e.~nearly two orders of magnitude less than the aforementioned example.
As a general feature, we note that the order of magnitude of $T_{\parallel}$ 
is settled by the energy scale of the problem:
all quantities involved in Eq.~\ref{eq:Tlong} are of the order of eV,
whose corresponding time scale is fs.
Therefore, the longitudinal relaxation of the spin considered here is extremely fast.
The physical reason is the large exchange splitting dominating the relaxation process,
which makes it energetically very expensive to modify the length of the moment
due to the high energies involved.

\subsubsection{Connection to experimental measurements}

Let us next consider the experimental 
scenario regarding the measurement of the
longitudinal spin-relaxation time.
For this, we first note that in a experiment,
several different mechanisms can contribute to this relaxation process,
whose overall relaxation time is generally denoted as $T_{1}$.
In this context, $T_{\parallel}$ calculated here is a particular
contribution to $T_{1}$, which may include further contributions depending on the
physical processes taking place. 
To the best of our knowledge, the first experimental 
technique that measures $T_{1}$  in magnetic single adatoms was developed by
Loth and co-workers~\cite{loth_measurement_2010}.
Within this STM-based technique, the spin-relaxation time was measured 
by monitoring the decay of electrons in excited states after the application of an 
all-electronic pump-probe scheme~\cite{loth_measurement_2010}.
It is noteworthy that this scheme has so far only been applied 
to adatoms deposited on semi-insulating substrates, which  are  
close to the atomic limit.
The original work by Loth and co-workers measured $T_{1}\sim90$ ns for a Fe-Cu dimer on
$\text{Cu}_{2}$Ni/Cu(100)~\cite{loth_measurement_2010}. 
A subsequent work by Rau and co-workers measured  $T_{1}\sim200$ $\mu$s
for a single Co atom on MgO/Ag(100)~\cite{rau_reaching_2014}.
Lastly, Baumann and co-workers reported 
$T_{1}\sim90$ $\mu$s for a single Fe atom on MgO/Ag(100)~\cite{baumann_electron_2015},
while in a recent work of Paul and co-workers on the same system,~\cite{paul_control_2017} 
the value of $T_{1}$ was enhanced up to the ms regime by fine-tuning
external conditions such as the height of the STM tip. 
To conclude, we note that the reported 
time resolution of the measuring technique employed in the above experiments 
ranges between few ns to hundreds of ps.

All the above measured values of the spin-relaxation time 
are several orders of magnitude larger 
than the values of order fs  that we discussed
in this review for $T_{\parallel}$ (see Fig.~\ref{fig:Tparallel}(a)).
Let us first point out that all of the above experiments were performed under
externally applied static magnetic fields that range between 1 T and 10 T. This, in turn,
breaks the degeneracy of the spin ground states~\cite{loth_measurement_2010},  
a situation that is commonly modeled by 
a shifted discrete energy diagram as the one shown in Fig.~\ref{fig:Tparallel}(c). 
We note that excitations within such a diagram 
are not allowed to change the length of the spin moment
(spin quantum number $S$ in this context), 
but only its projection (magnetic quantum number $S_{z}$).
Therefore, the main spin-relaxation process contributing to $T_{1}$
within such an scheme involves transitions between the two non-degenerate states
with same $S$ but opposite $S_{z}$ (see Fig.~\ref{fig:Tparallel}(c)).
We note that their energy separation is of order meV, hence much smaller than the 
excitations of order eV involved in the change of the spin magnetic moment size
considered for 
the calculation of $T_{\parallel}$, as schematically depicted in Fig.~\ref{fig:Tparallel}(b).
Furthermore, given that direct transitions between the two non-degenerate 
states of Fig.~\ref{fig:Tparallel}(c) are virtually nonexistent, 
spin-relaxation in these conditions is  driven by quantum tunneling processes,
which are intrinsically much slower than the direct transitions considered here. 
These two considerations explain why the spin-relaxation time  measured
under the mentioned experimental conditions
is several orders of magnitude larger than the values of $T_{\parallel}$ analyzed in this review.

It is apparent that, in order to experimentally
access the dynamics encoded into  $T_{\parallel}$, 
a measuring scheme based on ultrafast techniques that modify the
length of the spin magnetic moment is required. 
Considering the technological developments within 
STM measuring techniques~\cite{kruger_attosecond_2011,cocker_ultrafast_2013,cocker_tracking_2016}, 
accessing the fs time scale of magnetic adatoms seems to be a reasonable goal
for the near future by, e.g.~using ultrafast laser pulses, 
a breakthrough that would allow to monitor the ultrafast spin-dynamics
analyzed here.

\subsection{Transverse component}
\label{subsec:trans}

\begin{figure}[t]
\centering
\includegraphics[width=0.95\textwidth]{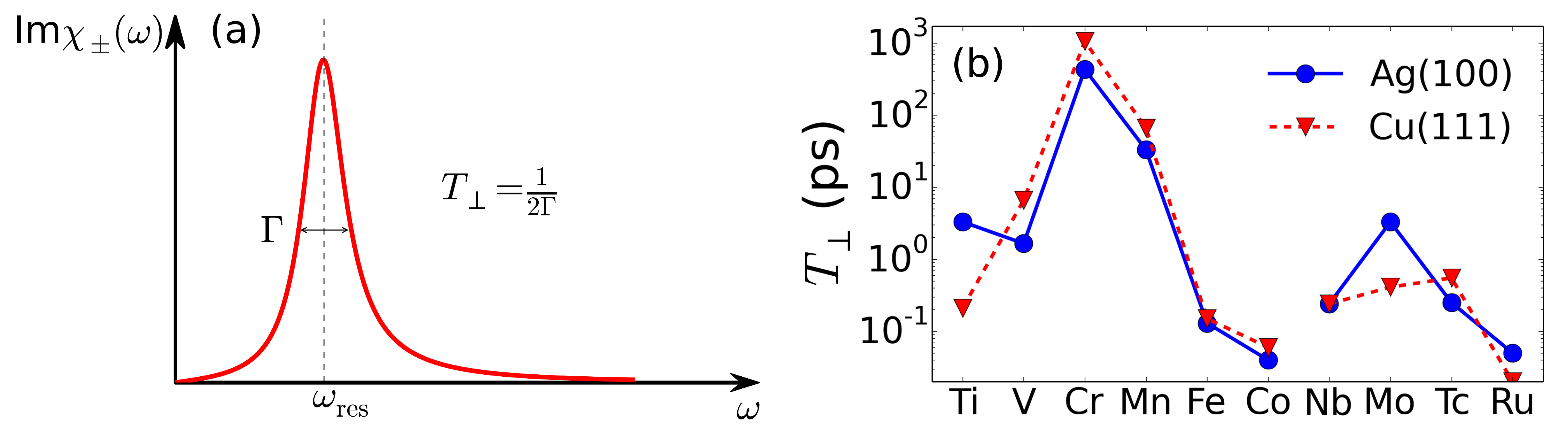}
\caption{Transverse relaxation time. (a) Schematic illustration of a transverse spin-excitation of a single adatom contained
in $\Im\chi_{\pm}(\omega)$. The resonance frequency, 
$\omega_{\text{res}}$, and the width, $\Gamma$, are indicated in the figure.
(b) Calculated values for the transverse 
relaxation time $T_{\perp}$ for magnetic 3\textit{d} and 4\textit{d} 
    adatoms on Ag(100) (blue circles) and Cu(111) (red triangles). 
    Note the logarithmic scale in the y axis.
}
\label{fig:t-trans}
\end{figure}

While longitudinal spin-susceptibility  contains 
excitations that modify the spin density, the transverse one describes 
damped precessional motion of the spin moment~\cite{white_quantum_2007}.
This motion, in turn, is described by 
the imaginary part of the enhanced spin-susceptibility,
$\Im\chi_{\pm}(\omega)$, 
a quantity that has been analyzed in depth in Sec.~\ref{sec:zpsf} (in particular, 
see Fig.~\ref{fig:imchi-xi}).
The characteristic form of the spin-excitation hosted by $\Im\chi_{\pm}(\omega)$
and its connection to the transverse spin-relaxation time is 
illustrated in Fig.~\ref{fig:t-trans}(a).
As discussed in Sec.~\ref{sec:zpsf}, $\Im\chi_{\pm}(\omega)$ is mainly determined by the 
resonance frequency, $\omega_{\text{res}}$, which is settled by the SOC,
and the width of the spin-excitation, $\Gamma$, which is proportional to the 
damping term $\eta$ (see in particular Eq.~\ref{eq:Gamma-eta-w0} in Appendix \ref{appendix:LLG}).
Noteworthily, the main contribution of the hybridization to $\Gamma$ is proportional
to the electron-hole excitations of opposite spin channel, 
$n_{\mathrm{e-h}}' = \pi\,\rho_{\mathrm{F},\uparrow}\,\rho_{\mathrm{F},\downarrow}$~\cite{PhysRevB.91.104420}:
\bek\label{eq:hybr-trans}
\Gamma\simeq \dfrac{n_{\mathrm{e-h}}'}{Q}\, \omega_{\text{res}},
\ek
with $Q=\mathrm{Re}\,\partial\chi^{\text{KS}}_{\pm}(\omega)/\partial\omega\big|_{\omega=0}$. 
We note that the order of magnitude of $\omega_{\text{res}}$ ranges between 
$10^{-2}-1$ meV while
$n_{\mathrm{e-h}}'/Q$ is a unitless fraction that is typically of order 
unity.

Importantly, a finite width corresponds to a finite 
damping of the precessing magnetic moment 
and is thus directly linked to
the transverse spin-relaxation time (see Appendix \ref{appendix:LLG}
and Eq.~\ref{eq:hybr-trans}): 
\bek\label{eq:Tperp-Gamma}
T_{\perp}=\dfrac{2}{\Gamma}\propto \left(n_{\mathrm{e-h}}'\right)^{-1}.
\ek
We note that, while $T_{\parallel}$ in Eq.~\ref{eq:Tlong}
is directly proportional to the density of 
spin-conserving electron-hole excitations $n_{\mathrm{e-h}}$,
$T_{\perp}$ above is inversely proportional to the 
spin-flip counterpart $n_{\mathrm{e-h}}'$.
Let us also note that an expression with similar physical implications has  been derived 
in the context of spin models with Kondo exchange in open quantum systems,
the so-called non-adiabatic contribution to the spin-decoherence rate~\cite{delgado_spin_2017}
(see in particular Eq.~(70) in that reference).

In Fig.~\ref{fig:t-trans}(b) we show calculated values for  $T_{\perp}$ 
extracted from the corresponding spin-excitation width (see Fig.~\ref{fig:imchi-xi}).
We note that the variation of $T_{\perp}$ among adatoms is considerably
larger (up to two orders more) than that of 
$T_{\parallel}$, shown in Fig.~\ref{fig:Tparallel}(a). 
This is a consequence of the large variation of the width $\Gamma$ 
of atomic spin-excitations, which can range from 10$^{-2}$ meV to 
few meV~\cite{dias_relativistic_2015,ibanez-azpiroz_zero-point_2016},
i.e.~nearly three orders of magnitude change.
This, in turn, can be linked to the electronic  DOS at the Fermi level via Eq.~\ref{eq:hybr-trans};
adatoms where the DOS peak of the \textit{d}-states lies close to the Fermi level, case of  
Ti, V, Fe and Co in Fig.~\ref{fig:M0}(c), tend to be much more hybridized than those 
where only the tail of the DOS peak lies at the Fermi level, case of Cr and Mn in Fig.~\ref{fig:M0}(c).
In this way, Cr and Mn acquire large relaxation times of  $T_{\perp}\sim 10^{1}$--$10^{2}$ ps,
while strongly hybridized adatoms such as Co, 
Nb and Ru
have $T_{\perp}\sim 10^{-1}$--$10^{-2}$ ps.
In all cases, the transverse spin-relaxation process is orders of magnitude slower than the longitudinal one, 
$T_{\perp}\gg T_{\parallel}$.

To conclude, let us note that the trend of $T_{\perp}$
within each \textit{d}-shell row is opposite to that shown by $T_{\parallel}$ (see Fig.~\ref{fig:Tparallel}(a)).
This comes as a consequence of the dependence on the density of electron-hole excitations,
with $T_{\parallel}\propto n_{\mathrm{e-h}}$ (see Eq.~\ref{eq:Tlong})
and $T_{\perp}\propto (n_{\mathrm{e-h}}')^{-1}$ (see Eq.~\ref{eq:hybr-trans}).
In fact, $T_{\parallel}$ and $T_{\perp}$
can be formally related to each other considering the relation between
the spin-conserving and spin-flip electron-hole contributions, namely
$\rho_{\mathrm{F}}^{2}/2=n_{\mathrm{e-h}}+n_{\mathrm{e-h}}'$. 
From Eq.~\ref{eq:Tlong} for $T_{\parallel}$ 
and Eqs.~\ref{eq:hybr-trans} and~\ref{eq:Tperp-Gamma} for $T_{\perp}$ one can then infer
the following expression:
\bek\label{eq:relation-Tpara-Tperp}
\dfrac{\rho_{\mathrm{F}}^{2}}{2}\simeq
T_{\parallel}\dfrac{U_{\parallel}\rho_{\mathrm{F}}-1}{U_{\parallel}}
+\frac{2}{T_{\perp}}\,\dfrac{Q}{\omega_{\text{res}}}.
\ek
The unifying concept behind the above relationship between $T_{\parallel}$ and $T_{\perp}$
is the hybridization of substrate electrons with the \textit{d}-states of the transition metal 
adatoms, which in essence gives rise to a finite total $\rho_{\mathrm{F}}$. 
However, despite the formal relationship,
the fact that 
$T_{\parallel}$ and $T_{\perp}$ in Eq.~\ref{eq:relation-Tpara-Tperp}
have fundamentally different prefactors
makes the time scale of the two relaxation constants
differ by nearly three orders of magnitude.

\subsubsection{Connection to experimental measurements}

Next, we consider several experimental measurements of 
spin-excitation lifetimes of different single adatoms and 
connect them to this work.
The lifetime of an atomic spin-excitation can be experimentally accessed  
from the width of the step observed in IETS $\text{d}I/\text{d}V$ measurements, which
provides a measure of $\Gamma$. 
Given that the energy resolution of this technique is 10$^{-1}$ meV at best~\cite{Khajetoorians2013},
the longest lifetimes that can be inferred following this procedure
are of order 10 ps (see Eq. \ref{eq:Tperp-Gamma}).  
This type of experiments can measure adatoms deposited on both 
metallic and semi-insulating substrates;
as a general trend, the latter induce a larger lifetime than the former 
due to a far smaller electronic hybridization.
We  begin by considering Ref.~\cite{Khajetoorians2011}, where 
Khajetoorians and co-workers estimate the spin-excitation lifetime  
of a Fe adatom deposited on metallic Cu(111) to be 
0.2 ps, in very good quantitative 
agreement with the calculated value $T_{\perp}=0.15$ ps for the same system
(see Fig.~\ref{fig:t-trans}(b)).
Noteworthily, when the same atom is deposited on metallic Pt(111), 
the measured lifetime  is increased by
more than an order of magnitude~\cite{Khajetoorians2013,hermenau_long_2018}.
Note that one finds a similar variation 
between the two substrates  considered here
for the elements Ti, V, Cr and Mo, as it can be checked in Fig.~\ref{fig:t-trans}(b). 
Focusing next on the semi-insulating $\text{Cu}_{2}$Ni/Cu(100) substrate,
a lower bound of  $\sim$10 ps has been experimentally estimated for
Fe~\cite{hirjibehedin_large_2007,loth_njp}, Mn~\cite{loth_njp} and Co~\cite{otte_role_2008}
adatoms, although it is possible that the actual lifetimes are substantially larger.
In fact, the calculations on Cr and Mn, which are the elements with smallest hybridization
and thus the ones closest to the semi-insulating limit, show that  $T_{\perp}$ can reach up to
10$^{3}$ ps (see Fig.~\ref{fig:t-trans}(b));
hence, it is not unlikely that the lifetimes of the aforementioned  
adatoms on $\text{Cu}_{2}$Ni/Cu(100) could be of the same order of magnitude.
Last, it is worth noting the case of Co on MgO~\cite{rau_reaching_2014},
which, despite being a semi-insulating substrate, yields a relatively short 
spin-excitation lifetime of $\sim$ 0.5 ps, i.e.~a common value for
adatoms deposited in metallic substrates analyzed in this review (see Fig.~\ref{fig:t-trans}(b)).


\section{Paramagnetic spin-excitations}
\label{sec:param}

In this last section, we switch the main subject of study from magnetic to nonmagnetic adatoms, 
i.e., adatoms where the Stoner criterion for magnetism is nearly fulfilled. 
In particular, we analyze their dynamical properties, which can  exhibit
fingerprints of magnetism in the form of well-defined 
features in the spin-excitation spectrum, 
i.e., paramagnetic spin-excitations (PSE). 
Noteworthily, these excitations can be viewed as
persistent spin-fluctuation modes that are activated by temperature and 
thus produce a measurable impact on 
properties such as specific heat or electron 
effective-mass enhancement~\cite{doniach_theory_1967,lonzarich_effect_1985}.
Thus, the same concept that leads to the destabilization of the magnetic moment
seen in Sec.~\ref{sec:zpsf} can in this case instead be used to generate a signal of magnetic origin.

As in previous sections, the  central quantity 
for the discussion is the spin-excitation spectrum, which in this case is encoded into the 
paramagnetic spin-susceptibility 
(see Eq. \ref{eq:susc-long}):~\cite{aguayo_why_2004}
\begin{equation}\label{eq:chi-general}
\chi(\omega)=\dfrac{\chi^{\mathrm{KS}}(\omega)}{1-I_{\mathrm{s}}\,\chi^{\mathrm{KS}}(\omega)}.
\end{equation}
Above, $I_{\mathrm{s}}$ denotes the so-called Stoner parameter,
which plays the role of the exchange-correlation kernel in the 
adiabatic local spin-density approximation.
It is noteworthy that the static limit of Eq.~\ref{eq:chi-general} 
recovers a local form of the familiar Stoner theory that provides the well-known criterion
for magnetism, i.e. 
\bek\label{eq:stoner-crit}
\dfrac{\partial^{2}E(m)}{\partial m^{2}}\Big|_{m=0} = 
\chi(0)^{-1}<0\Rightarrow I_{\mathrm{s}}\,\rho_{\mathrm{F}}>1,
\ek
where we used $\chi^{\mathrm{KS}}(0)=\rho_{\mathrm{F}}$~\cite{aguayo_why_2004}.
In essence, the product $I_{\mathrm{s}}\,\rho_{\mathrm{F}}$ quantifies the competition
between the exchange interaction, which  
enhances the tendency towards magnetism of electrons in localized orbitals, 
and substrate hybridization, which induces delocalization of the adatom's electrons and 
therefore acts against magnetism, thus playing the role of 
the kinetic energy in the standard Stoner theory.
It is interesting to note that 
even if an adatom does not fulfill the Stoner criterion,
it can still develop dynamical PSE provided the details of the electronic structure make the denominator
of Eq.~\ref{eq:chi-general} vanishingly small at a finite frequency.

\begin{figure}[t]
\centering
\includegraphics[width=0.95\textwidth]{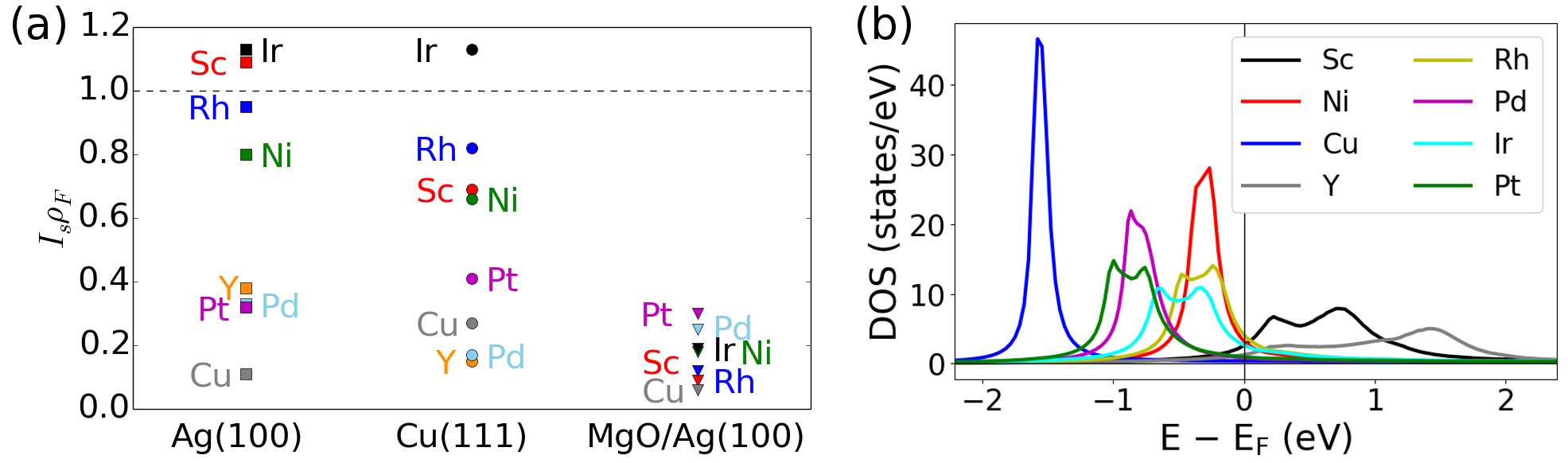}
\caption{Ground state electronic properties of various 3\textit{d}, 4\textit{d} and 5\textit{d}
transition metal adatoms. 
(a) Calculated Stoner product for adatoms deposited on Ag(100) (squares), Cu(111) (circles) and MgO/Ag(100) (triangles).
(b) Calculated non-magnetic DOS of adatoms on Ag(100). The vertical line denotes the Fermi level.
}
\label{fig:IS-table}
\end{figure}

\subsection{Searching for the optimal candidates via the Stoner criterion}

We begin by characterizing the set of 3\textit{d}, 4\textit{d} and 5\textit{d} 
transition metal adatoms
that could potentially exhibit PSE. 
For this purpose, we consider the Stoner criterion of 
Eq.~\ref{eq:stoner-crit} and search for adatoms whose Stoner product is close to 1.
In Fig.~\ref{fig:IS-table}(a)
we plot $I_{\mathrm{s}}\,\rho_{\mathrm{F}}$ for 
several adatoms deposited on the Ag(100) and Cu(111) substrates analyzed throughout the review.
In addition, we have also included the MgO/Ag(100) substrate in order to 
analyze the differences introduces by a semi-insulating layer.
We first note that the Stoner products of all the adatoms 
are below or slightly above 1.
As a general trend, the calculations show that the metallic substrates Ag(100) and Cu(111) host adatoms
whose Stoner product is closer to the critical value 1 as compared to semi-insulating MgO/Ag(100).
This is mainly due to the small $\rho_{\mathrm{F}}$ in the latter,
as tabulated in Table~\ref{table:stoner}.
Among the two metallic substrates, Ag(100)  
hosts adatoms whose Stoner product are closest to 1, 
with $I_{\mathrm{s}}\,\rho_{\mathrm{F}}$ ranging between $\sim[1-0.1,1+0.1]$ for Sc, Ir, Rh and Ni adatoms.
Therefore, in the rest of this Section we focus on discussing the Ag(100) substrate in detail,
as it illustrates best the concept of PSE.

In Fig.~\ref{fig:IS-table}(b) we show the calculated DOS of the considered 3\textit{d}, 4\textit{d} and 5\textit{d} 
adatoms. The figure shows that the large peaks coming from localized \textit{d}-states
acquire a finite width, which ranges from $\sim$0.2 eV in the case of Cu, up to
$\sim$1 eV in the case of Sc and Y. This feature reveals a strong 
hybridization of the adatom's \textit{d}-states with the electron bath of the substrate~\cite{lounis_theory_2011,PhysRevB.91.104420}.
As a consequence, the DOS acquires a finite value at the Fermi level, $\rho_{\mathrm{F}}$, which is
listed in Table~\ref{table:stoner} for all adatoms, 
alongside with the calculated Stoner parameter and the resulting Stoner product.
We note that the largest $\rho_{\mathrm{F}}$ corresponds to Rh due to the 
large width of the nearly filled
\textit{d}-peak in the DOS and its proximity to the Fermi level (see Fig.~\ref{fig:IS-table}(b)).
It is also noteworthy that Sc and Ir, which are predicted to be slightly magnetic, 
have a smaller $\rho_{\mathrm{F}}$ than Rh and Ni. This is compensated by the Stoner parameter, which
is almost 50\% larger as compared to Ni and Rh. In this respect, the case of Ni
is specially interesting, given that the calculated $I_{\mathrm{s}}$ for the adatom
is almost 50\% smaller compared to the bulk value~\cite{janak_uniform_1977}.
This drop of $I_{\mathrm{s}}$ is the reason why Ni is magnetic in bulk
but non-magnetic as an adatom on Ag(100) 
and other metallic substrates~\cite{PhysRevB.65.104441,beckmann_magnetism_1997}.

\begin{table}[t]
\centering
 \begin{tabular}{| c | c | c | c | c | c | c | c | c |}
  \hline                       
                           & Sc & Ni & Cu & Y & Rh & Pd & Ir & Pt  \\ 
  \hline  
   $\rho_{\mathrm{F}}$ (states/eV)  & 2.65 & 3.04 & 0.35 & 1.37 & 3.51 & 0.77 & 2.81 & 0.89 \\
   \hline  
 $I_{\mathrm{s}}$ (eV)              & 0.41 & 0.29 & 0.32 & 0.28 & 0.27 & 0.39 &  0.40 & 0.36 \\  
         \hline  
  $I_{\mathrm{s}}\,\rho_{\mathrm{F}}$      & 1.09 & 0.88 & 0.11 & 0.38 & 0.95 & 0.33 &  1.12 & 0.32 \\  
         \hline  
    \end{tabular}
     \caption{Calculated values for the DOS at the Fermi level ($\rho_{\mathrm{F}}$),
       Stoner parameter ($I_{\mathrm{s}}$) and Stoner product ($I_{\mathrm{s}}\,\rho_{\mathrm{F}}$) for  several 3\textit{d}, 4\textit{d} and 5\textit{d} 
adatoms on Ag(100).\label{table:stoner}}
        \end{table}

\begin{figure}[t]
\centering
\includegraphics[width=0.95\textwidth]{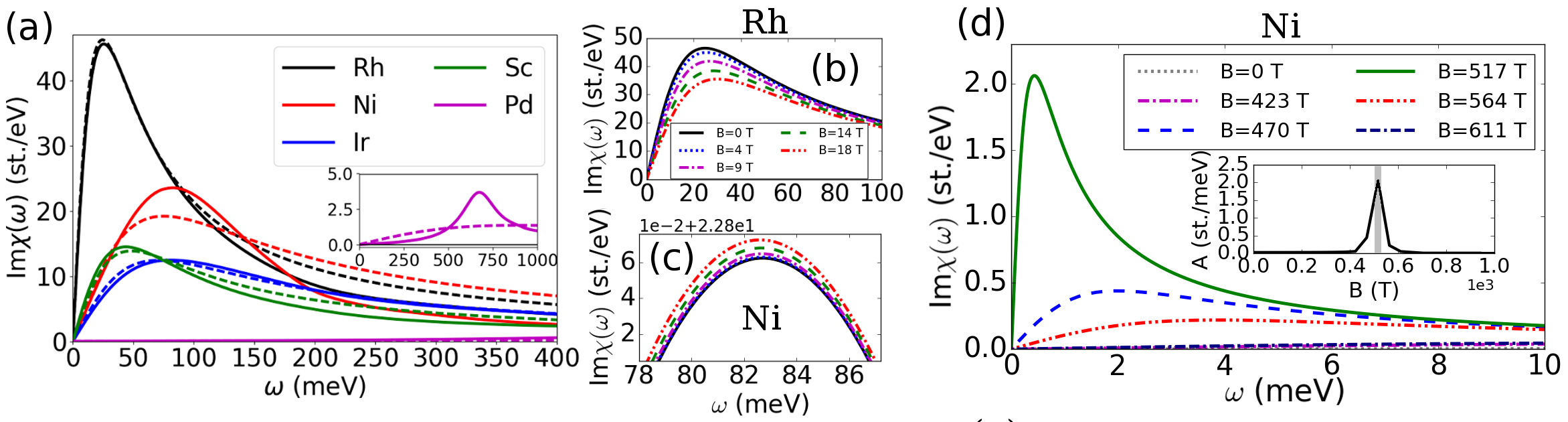}
\caption{Spin-excitation spectrum of selected adatoms.
(a) Solid lines illustrate the calculated density of PSE as given by $\Im\chi(\omega)$  (Eq.~\ref{eq:chi-general}) of selected 3\textit{d}, 4\textit{d} and 5\textit{d} transition metal adatoms deposited on the metallic Ag(100) substrate. 
Dashed lines denote the approximation of Eq.~\ref{eq:imchi-linear-w}.
Note that both Eqs.~\ref{eq:chi-general} and \ref{eq:imchi-linear-w} give rise to PSE.
(b) and (c) Density of PSE as given by $\Im\chi(\omega)$ for Rh and Ni adatoms, respectively, for magnetic fields of up to 18 T (both figures share the same legend).
(d) Same as in (c) but for larger magnetic fields of up to 10$^{3}$ T.
The inset depicts the evolution of the PSE's amplitude (see Eq.~\ref{eq:resonance-freq}) as a function of the magnetic field.
}
\label{fig:PSE}
\end{figure}

\subsection{Fingerprints of magnetism in the dynamical regime}

In Fig.~\ref{fig:PSE}(a) we illustrate the  calculated spin-excitation spectra as given by 
$\Im\chi(\omega)$ from Eq.~\ref{eq:chi-general},
where all calculations were done considering the non-magnetic ground state.  
Interestingly, Fig.~\ref{fig:PSE}(a) reveals peak-like structures resonating at frequencies below 100 meV
for Rh, Ni, Ir and Sc adatoms.
This is exceptional, as most non-magnetic elements 
exhibit a featureless spectrum owing to a complete overdamping of the excitations.
Rh represents the most favorable case, displaying
a well-defined peak at $\omega_{\text{res}}\sim 20$ meV 
and a width of 
$\Delta\sim50$ meV,
the associated lifetime being $\sim30$ fs.
It is noteworthy that both the lifetime and the height of the peak, 
the latter being related to the intensity of the excitation, are only one order
of magnitude smaller than those of usual transverse spin-excitations
of magnetic adatoms studied in Sec. \ref{sec:zpsf} (see in particular Fig. \ref{fig:imchi-xi})
and that are accessible by IETS 
(see, e.g., Fe on Cu(111) in Refs.~\cite{Khajetoorians2011,dias_relativistic_2015}).
On the other extreme, Pd in Fig.~\ref{fig:PSE}(a) shows a highly overdamped resonance at around
600 meV (see figure inset) whose intensity is an order of magnitude smaller than that of Rh.
Therefore, the  calculations reveal the existence of PSE 
whose resonance frequency and width vary strongly depending on the adatom.

Next, we focus on characterizing the physical mechanism behind PSE
that allows an interpretation of the \textit{ab initio} results displayed in Fig.~\ref{fig:PSE}(a).
For this purpose, let us consider the frequency expansion of the paramagnetic KS spin response function
up to linear order, i.e., 
$\chi^{\mathrm{KS}}(\omega)=\rho_{\mathrm{F}}+\mathrm{i}\,\alpha\,\omega + \mathcal{O}(\omega^{2})$.
One can show~\cite{PhysRevLett.119.017203} that the linear expansion
coefficient is well approximated by $\alpha\sim -\pi \rho^{2}_{\mathrm{F}}/4$.
Therefore, the spin-excitation spectrum within this approximation 
is given by a simple expression involving only the DOS at $E_{\mathrm{F}}$ and the Stoner parameter:
\begin{equation}
\label{eq:imchi-linear-w}
\Im\chi(\omega)= \dfrac{\pi}{4}\dfrac{\rho^{2}_{\mathrm{F}}\,\omega}{\big(1-I_{\mathrm{s}}\,\rho_{\mathrm{F}} \big)^{2}
+(\frac{\pi}{4}I_{\mathrm{s}}\rho^{2}_{\mathrm{F}}\,\omega)^{2}}.
\end{equation}
The above expression can be readily computed by using the calculated values for  $\rho_{\mathrm{F}}$ and $I_{\mathrm{s}}$ shown in Table \ref{table:stoner}: 
the results are displayed in Fig.~\ref{fig:PSE}(a) (see dashed lines). 
A comparison to the full \textit{ab initio} calculations (solid lines)
reveals a very good agreement for frequencies below 100 meV 
in the case of Rh, Ir and Sc, where both the peak and width are 
properly described within $\leqslant 10\%$ relative error.
This error is considerably larger in the case of Ni, indicating the importance
of higher order expansion terms in $\omega$ for this case. 
Finally, the peak for Pd is far beyond the limit of small frequencies and therefore
the approximation of Eq.~\ref{eq:imchi-linear-w} breaks down.

\begin{figure}[t]
\centering
\includegraphics[width=0.5\linewidth]{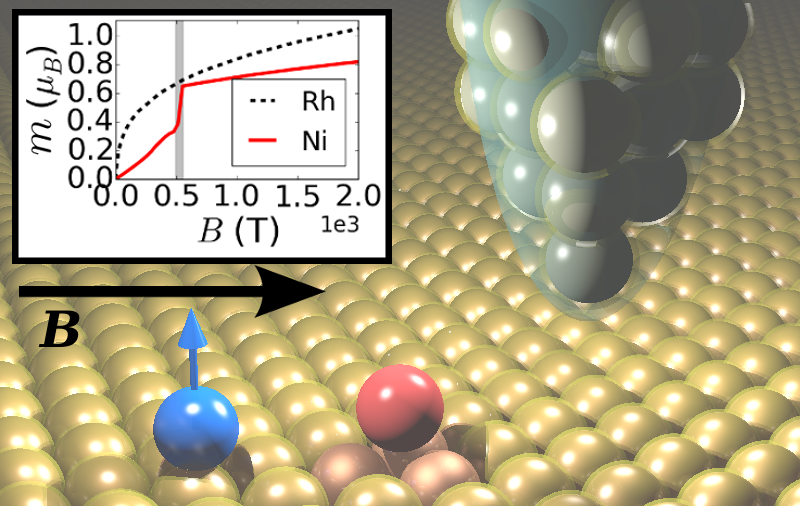}
\caption{ Minimal setup illustrating the proposed IETS measurement. 
Substrate, non-magnetic adatom and tip atoms are displayed as 
gold, red and grey balls, respectively, while the black arrow
depicts an external magnetic field $B$.
The graph in the inset illustrates the calculated magnetic moment $m$ as a function
of the external field for Rh and Ni adatoms, with the grey area indicating the
critical regime of Ni where $m$ shows a discontinuity. 
A blue ball with an arrow has been added in the main figure to illustrate the possibility of 
coupling a magnetic adatom to the non-magnetic one, inducing on the latter a
magnetic moment of the order of the values shown in the 
inset, thus mimicking the effect of large magnetic fields~\cite{footnote}.
}
\label{fig:STM}
\end{figure}

Proving Eq.~\ref{eq:imchi-linear-w} to be an accurate approximation 
of the full spin-excitation density given by Eq.~\ref{eq:chi-general} is extremely convenient, 
as the former provides an analytical interpretation for the origin of PSE 
in terms of just $\rho_{\mathrm{F}}$ and $I_{\mathrm{s}}$, two basic electronic properties of adatoms.
Indeed, the resonance frequency, linewidth and 
amplitude of PSE 
predicted by Eq.~\ref{eq:imchi-linear-w} can be cast into simple expressions:
\begin{equation}
\label{eq:resonance-freq}
\begin{aligned}
&\omega_{\text{res}}=\dfrac{4}{\pi}\frac{|1-I_{\mathrm{s}}\,\rho_{\mathrm{F}}|}{I_{\mathrm{s}}\,\rho^{2}_{\mathrm{F}}},\;\; 
\Delta = 2\sqrt{3}\,\omega_{\text{res}},\\
& \;\; A\equiv \Im\chi(\omega_{\text{res}})=\dfrac{1}{2I_{\mathrm{s}}|1-I_{\mathrm{s}}\,\rho_{\mathrm{F}}|} .
\end{aligned}
\end{equation}
Interestingly, a potential  measurement  of the above quantities
would directly yield experimental estimates for $\rho_{\mathrm{F}}$ and $I_{\mathrm{s}}$.
In closer inspection, one recognizes the Stoner product $I_{\mathrm{s}}\,\rho_{\mathrm{F}}$ 
as the key quantity in Eq.~\ref{eq:resonance-freq}; 
as $I_{\mathrm{s}}\,\rho_{\mathrm{F}}\rightarrow 1$ (i.e., magnetic instability),
the resonance frequency as well as the linewidth tend 
to zero while the intensity of PSE
shows a singularity. 
This analysis offers therefore the interpretation we seeked for, namely 
that elements closer to the magnetic instability 
show enhanced PSE, as it can be clearly checked from the comparison of
Figs.~\ref{fig:IS-table}(a) and \ref{fig:PSE}(a). 
We emphasize that the mechanism just described is 
fundamentally different from the one taking place in the magnetic adatoms
analyzed in the previous sections, 
where the resonance frequency of transverse spin-excitations 
is settled by the SOC via the MAE.

\subsection{Potential means of measurement: connection to dI/dV curves}

We next focus on assessing the potential impact of PSE on the
$\mathrm{d}I/\mathrm{d}V$ signal as measured in IETS experiments,
the technique of choice for measuring magnetic excitations 
(see, e.g., 
Refs.~\cite{Khajetoorians2011,Khajetoorians2013,steinbrecher_absence_2016}).
The corresponding minimal setup is illustrated in Fig.~\ref{fig:STM},
which displays an STM tip 
measuring the adatom's excitations under an applied external magnetic field, denoted
as $B$. 
We first notice that
PSE respond to magnetic fields by shifting their resonance frequency.
This is quantitatively demonstrated in Figs.~\ref{fig:PSE}(b) and \ref{fig:PSE}(c), 
where the calculated spin-excitation spectra
are shown for Rh and Ni adatoms, respectively, 
under $B$ fields of $\sim$10 T that are achieveable in state-of-the-art laboratories
(see, e.g., Refs.~\cite{Khajetoorians2013,rau_reaching_2014,donati_magnetic_2016}). 
Noteworthily, while the PSE of Rh 
shifts towards larger frequencies as $B$ is increased (see Fig.~\ref{fig:PSE}(b)),
the PSE of Ni exhibits the opposite behavior (see Fig.~\ref{fig:PSE}(c)).
This difference arises from the fact that magnetic fields 
induce an effective modification of Stoner product, i.e.,
$I_{\mathrm{s}}\, \rho_{\mathrm{F}}\rightarrow \xi(B)\,I_{\mathrm{s}}\, \rho_{\mathrm{F}}$,
where $\xi(B)$ is a term that depends both on the magnetic field 
as well as on the adatom's electronic structure~\cite{PhysRevLett.119.017203}.
In particular, the details of the latter make $\xi(B)>1$ for Ni while 
$\xi(B)<1$ for Rh~\cite{PhysRevLett.119.017203}, leading to the aforementioned divergent responses
in accordance with Eq.~\ref{eq:resonance-freq}.

Remarkably, when strong enough magnetic fields are applied to Ni, 
the modified Stoner criterion
can be tuned towards the critical point, 
as shown in Fig.~\ref{fig:PSE}(d).
As a consequence, the PSE's resonance frequency approaches the origin in a singular way
while the amplitude of the excitation is enhanced by as much as two orders of magnitude for 
$B\sim 500$ T.  
It is interesting to note that this critical behavior 
is also present on the $B$-field dependence of the induced magnetic moment $m$, 
as shown in the inset of Fig.~\ref{fig:STM}.
While Rh shows a continuous dependence, Ni reveals a discontinous transition 
at approximately the critical field value $B\sim 500$ T, above which
the system enters a magnetic regime where 
the internal exchange field effectively contributes to  $m$ on top of the
external Zeeman field, featuring the atomic version of
a quantum phase transition.
We note that, although such large $B$ fields 
are clearly out of reach for current experiments,
this feature could be potentially observed,  
e.g., 
via the proximity effect, by placing a 
magnetic adatom in the neighborhood of the non-magnetic one
(see Fig.~\ref{fig:STM}).
Calculations show that the former can induce on the latter a magnetic moment of the
same order of magnitude as the one induced by the fields 
of Fig.~\ref{fig:PSE}(d)~\cite{footnote},
thus mimicking the action of large magnetic fields.

\begin{figure}[t]
\centering
\includegraphics[width=0.95\linewidth]{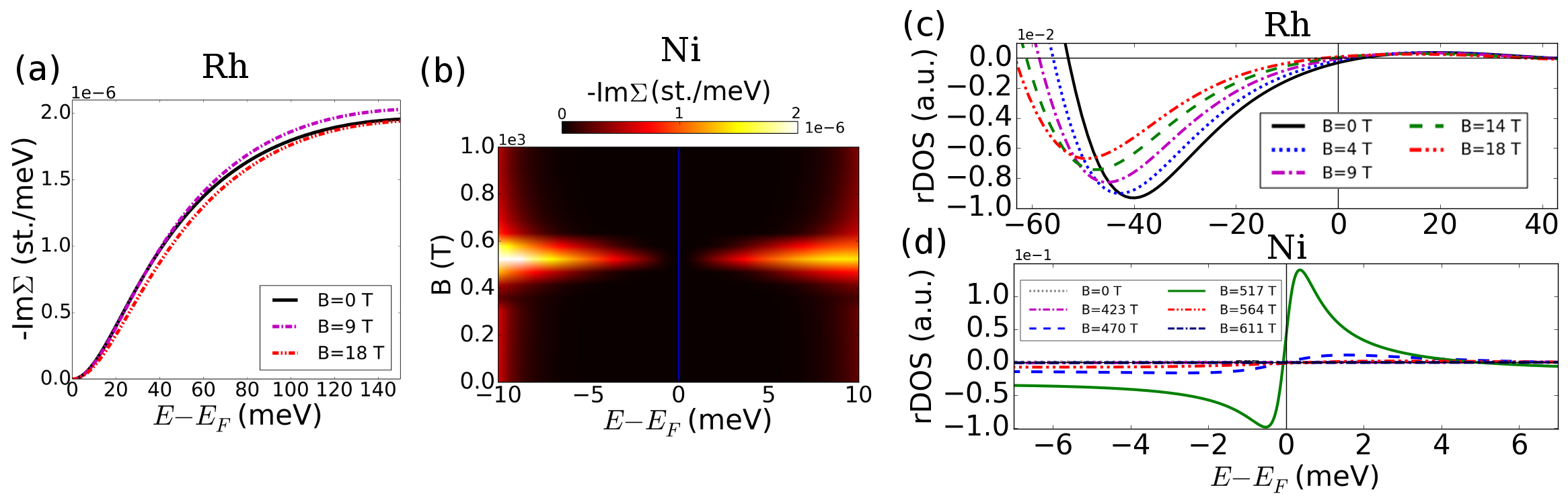}
\caption{Calculated electronic renormalization. 
(a) and (b) Imaginary part of the 
self-energy, $\Im\Sigma(V_{\mathrm{F}})$, for
Rh and Ni adatoms, respectively. Note the different 
scale of the magnetic field and the energy window in the two cases.
Vertical (blue) line in (b) separates negative and positive energies.
(c) and (d) Energy derivative of the renormalized DOS (s orbital)
for Rh and Ni, respectively.
Note the difference in magnitude on the applied magnetic fields in both cases.
}
\label{fig:selfe}
\end{figure}

Next we evaluate the impact of PSE on the $\mathrm{d}I/\mathrm{d}V$ signal
of an IETS measurement. For such purpose  we consider the so-called Tersoff-Hamann
approximation~\cite{PhysRevLett.50.1998,PhysRevLett.86.4132}, which relates the tunneling conductance to 
the electronic DOS at the tip position, which is renormalized by the adatom's excitations. 
The latter quantity can be accessed by means of a technique
that combines many-body perturbation theory with the TDDFT scheme; details
can be found in Ref.~\cite{Schweflinghaus2014}.
The central object within this formalism is the electron self-energy, 
$\Sigma$, which contains 
the interactions between the tunneling electrons from the tip at bias voltage $V$
and the adatom's PSE. It is particularly revealing
to inspect its imaginary part~\cite{Schweflinghaus2014},  
\begin{equation}
\label{eq:selfe}
\begin{aligned}
&\Im\Sigma(V_{\mathrm{F}}) = - I_{s}^{2}\int_{0}^{-V} \!\!\!\!\!\!\mathrm{d}\omega \; \rho(V_{\mathrm{F}}+\omega) \Im\chi(\omega),
\end{aligned}
\end{equation}
with $\rho(E)$ the energy-dependent DOS, $V_{\mathrm{F}}=E_{\mathrm{F}}+V$ and $E_{\mathrm{F}}$
the Fermi energy.  The calculated $\Im\Sigma(V_{\mathrm{F}})$  is shown in Fig.~\ref{fig:selfe}(a) 
for Rh under various magnetic fields of up to 18 T.  
The results reveal a clear step for positive bias voltage
that saturates at $\sim 100$ meV, i.e., after the PSE peak has been integrated 
(see Eq.~\ref{eq:selfe}). 
Note also that the calculated self-energy
slightly varies as a function of the magnetic field.
When larger magnetic fields are applied, as illustrated in Fig.~\ref{fig:selfe}(b) for the case of Ni,
the critical behavior of the PSE (see Fig.~\ref{fig:PSE}(d)) translates into  a
clear maximum at the value of the critical field,
where $\Im\Sigma(V_{\mathrm{F}})$ increases by an order of magnitude.

The presence of PSE has a broad effect on the 
renormalization of the DOS at the vacuum, where
IETS tips measure the signal. In particular, the energy derivative of the 
renormalized DOS (rDOS) is a quantity that is linked to the $\mathrm{d}^{2}I/\mathrm{d}V^{2}$ curve
measured by IETS~\cite{Schweflinghaus2014}.  
The former quantity is displayed in Fig.~\ref{fig:selfe}(c) for Rh, 
where the magnetic field dependence is clearly visible. 
Noteworthily, the calculations demonstrate that 
the tunneling electrons from the tip are able to trigger the PSE,
leading to a peak in the meV region that, furthermore, 
reacts to external magnetic fields
by shifting its resonance frequency as well as 
substantially modifying its intensity. 
We also note the strong asymmetric distribution 
between positive and negative frequencies, a feature that  emerges from the
background electronic structure~\cite{Schweflinghaus2014} and is 
commonly present in  $\mathrm{d}^{2}I/\mathrm{d}V^{2}$ curves measured on magnetic adatoms  
(see, e.g., 
Refs.~\cite{heinrich_single-atom_2004,Khajetoorians2011,PhysRevLett.110.157206,Khajetoorians2013,PhysRevLett.115.237202}).
On the other hand, when Ni is driven into the critical regime as in Fig.~\ref{fig:selfe}(d),
the calculations reveal a huge change of the signal's intensity 
as the PSE approaches the critical point. 
This analysis therefore shows that magnetism offers 
a prime way of manipulating PSE, enabling to discern them from other
excitations of similar energy but non-magnetic origin, such as phonons.

\section{Outlook}
\label{sec:conclusions}

In this review, we have presented a state-of-the-art treatment based on TDDFT 
of several dynamical aspects of single-adatoms deposited on metallic substrates.
As the first main topic, we have shown that the magnetic stability defined by the MAE can be
severely reduced as a consequence of quantum spin-fluctuations that arise
mainly from Stoner spin-flip excitations. 
Our analysis has quantitatively shown how protection against fluctuations is gained
by reducing the hybridization with substrate electrons as well as maximizing the adatom's local
magnetic moment. 
In this context, future work should address an important 
contribution that has not been considered here, namely the 
vibrational contribution of phonons,
which is believed to be key in systems of great importance like Ho on MgO/Ag(100)~\cite{donati_magnetic_2016}.
In particular, a reliable \textit{ab initio} scheme for the calculation of the electron-phonon 
coupling~\cite{Giustino2017,PhysRevLett.88.066805} and its  
effect on the magnetic stability of single-adatoms is necessary in order to
complement the knowledge gained on the electron-electron side.
A second important ingredient that should be further considered in the future from an \textit{ab initio} viewpoint
is the destabilizing effect of external particles, such as the tunneling electrons coming from an STM tip
or X-ray photons of an XMCD measurement. In both cases, 
it is experimentally clear that the magnetic stability and spin lifetimes are
severely affected by the intensity or flux of the incoming particles 
(see e.g., Refs~\cite{donati_magnetic_2016,natterer_reading_2017,paul_control_2017}.)
First steps towards incorporating such effects into an 
\textit{ab initio} scheme have been already taken in Ref.~\cite{Schweflinghaus2014}:
pending issues include the incorporation of SOC into the formalism as well as establishing a direct
connection between the renormalization induced by the tunneling 
electrons and the magnetic stability of the
adatom. 

The conclusions drawn from the 
single-adatom case  serve also as a solid platform for tackling 
more complex nanomagnets composed of several 
exchange-coupled magnetic atoms.
This playground is particularly attractive given that 
several STM experiments have reported 
clear observations of 
stable magnetic moments in this type of composite system
(see e.g., 
Refs.~\cite{loth_bistability_2012,khajetoorians_current-driven_2013,hermenau_gateway_2017}).
This strongly suggests that local fluctuation effects decrease with increasing cluster size,
a feature that is in qualitative 
accordance with one main finding presented throughout the review, 
namely that the relative impact of the fluctuations
becomes smaller the larger the magnetic moment gets.
This general trend, however, is likely to be modified by   
characteristics associated to clusters,
such as the appearance of optical modes in the spin-excitation spectrum 
(i.e. additional peaks in $\Im \chi_{\perp}(\omega)$, 
see e.g., Refs.~\cite{lounis_dynamical_2010,lounis_theory_2011}) 
that would contribute to the fluctuations as
seen from Eq.~\ref{eq:fd-theorem}.
On top of this, 
the impact of the fluctuations on the magnetic interactions among different adatoms
remains to be fully explored and clarified.

Throughout the review we have also paid close attention to 
the time-scales of different spin-relaxation processes. 
We have studied two particular contributions, 
namely the longitudinal and transversal 
spin-relaxation times, 
providing their connection to the electronic structure of single-adatoms
and, when possible, comparing the \textit{ab initio} data to experimental measurements. 
In the case of $T_{\perp}$, we have shown that it is governed by atomic spin-excitations 
that induce the precessional motion of the magnetic moment, hence this relaxation
time contributes to $T_{2}$ as commonly defined 
in the standard literature~\cite{white_quantum_2007,canet_relaxation_2006}.
Interestingly, there is yet another contribution to $T_{2}$ that 
has recently attracted a great interest in the 
single-adatom community, namely the adiabatic spin-decoherence time (see e.g., the review~\cite{delgado_spin_2017}).
In fact, very recent EPR-based experiments 
have provided hard numbers for this quantity~\cite{baumann_electron_2015,willke_probing_2018},
which is key for the potential use of single-adatoms as quantum devices.
The \textit{ab initio} calculation of this quantity is a challenging task for the future
due to its pure quantum origin, which is not trivial to capture within  the mean-field approach of DFT.
Regarding the longitudinal spin-relaxation time, 
we have argued  that,
although currently available techniques
cannot monitor the femtosecond regime of magnetic single adatoms, 
it is reasonable that this can be achieved in the near future, 
e.g., by employing STM-integrated 
ultrafast laser schemes~\cite{kruger_attosecond_2011,cocker_ultrafast_2013,cocker_tracking_2016},
thus giving access to the ultrafast spin-dynamics
described in this work.

As the last main topic, we have analyzed paramagnetic spin-excitations, 
i.e. fingerprints of magnetism in nonmagnetic adatoms, 
as well as means of detecting them in IETS experiments. 
Remarkably, a very recent measurement~\cite{PhysRevB.96.035155} of
the conductance associated to a single Pd adatom deposited
on Pd(111) has been  interpreted as being strongly affected by paramagnon scattering,
i.e., a potential indirect experimental observation of PSE.
This opens up a promising range of applications  
for non-magnetic adatoms in nanotechnology, 
which encodes and manipulates information  
into excitation modes like PSE.

\section*{Acknowledgments}

We gratefully acknowledge discussions 
and common work with B. Schweflinghaus, 
as well as fruitful scientific exchange with F. Guimar\~aes,
A. Eiguren, N. Lorente and P. H. Dederichs on the theory side, and J. Hermenau
and J. Wiebe on the experimental side. 
This work has been supported by the
Helmholtz Gemeinschaft Deutscher-Young Investigators
Group Program No. VH-NG-717 (Functional Nanoscale
Structure and Probe Simulation Laboratory), the Impuls und
Vernetzungsfonds der Helmholtz-Gemeinschaft Postdoc Programme,
and funding from the European Research Council (ERC) under
the European Union's Horizon 2020 research and innovation
programme (ERC-consolidator grant 681405 — DYNASORE).
We also acknowledge the computing
time granted by the JARA-HPC Vergabegremium and provided
on the JARA-HPC Partition part of the supercomputer JURECA at Forschungszentrum J\"ulich.

\appendix

\section*{Appendices}

\section{Longitudinal Bloch equation}
\label{appendix:Bloch-long}

The Bloch equation for the longitudinal change 
of the magnetization under the effect
of a time-dependent perturbation $H_{1}(t)$ along the longitudinal direction 
can be written as~\cite{white_quantum_2007}
\begin{equation}
\label{eq:bloch-T1-t}
\dfrac{\mathrm{d}m_{z}(t)}{\mathrm{d}t}=\dfrac{\chi^{\text{Bl}}_{0}H_{1}(t)-m_{z}(t)}{T_{\parallel}},
\end{equation}
with $\chi^{\text{Bl}}_{0}$ a static spin-susceptibility.
The above equation describes how $m_{z}(t)$ comes back to equilibrium 
with a characteristic relaxation time $T_{\parallel}$
after being perturbed by $H_{1}(t)$. 
Using $f(t)=\int\mathrm{d}\omega f(\omega)e^{-\mathrm{i}\omega t}$
for both $m_{z}(t)$ and $H_{1}(t)$ we can write Eq.~\ref{eq:bloch-T1-t} in the frequency domain,
\begin{equation}
\begin{split}
\label{eq:bloch-T1-w}
&m_{z}(\omega)\left(-\mathrm{i}\,\omega T_{\parallel} + 1\right)=\chi_{0}H_{1}(\omega) \Rightarrow \\
&
\dfrac{m_{z}(\omega)}{H_{1}(\omega)}\equiv \chi^{\text{Bl}}(\omega) =
\dfrac{\chi^{\text{Bl}}_{0}}{1-\mathrm{i}\,\omega T_{\parallel}},
\end{split}
\end{equation}
where  $\chi^{\text{Bl}}(\omega)$ is the enhanced spin-susceptibility. 
The real and imaginary parts of the above equation read
\begin{eqnarray}
\label{eqs:bloch-re}
& \Re \chi^{\text{Bl}}(\omega) =  \dfrac{\chi^{\text{Bl}}_{0}}{1+(\omega T_{\parallel})^{2}}, \\
\label{eqs:bloch-im}
& \Im \chi^{\text{Bl}}(\omega) =  \dfrac{\chi^{\text{Bl}}_{0}\omega T_{\parallel}}{1+(\omega T_{\parallel})^{2}}. 
\end{eqnarray}

Next, we consider the Taylor expansion of the KS spin-susceptibility
(see Eq.~\ref{eq:chi-KS-long} in the main text),
\bek\label{eq:chi-KS-long-appendix}
\chi^{\mathrm{KS}}_{\parallel}(\omega)\simeq \rho_{\mathrm{F}} - \mathrm{i}\,n_{\mathrm{e-h}}\omega.
\ek
The first-order expansion coefficient 
$n_{\mathrm{e-h}}=\pi(\rho^{2}_{\mathrm{F},\uparrow}+\rho^{2}_{\mathrm{F},\downarrow})/2$ has been calculated in the Supplemental
Material of Ref.~\cite{PhysRevLett.119.017203} .
Inserting $\chi^{\mathrm{KS}}_{\parallel}(\omega)$ of 
Eq.~\ref{eq:chi-KS-long-appendix} into the definition of the
TDDFT enhanced spin-susceptibility $\chi(\omega)$ 
(see Eq.~\ref{eq:susc-long} of the main text), the imaginary part
$\text{Im}\chi_{\parallel}(\omega)$ can be cast in the following way,
\begin{equation}
\begin{split}
\label{eq:susc-T1-conn}
 \Im\chi_{\parallel}(\omega) &=\dfrac{n_{\mathrm{e-h}}\omega}
{(1-U_{\parallel}\,\rho_{\mathrm{F}})^{2}}
\,\dfrac{1}
{1+\left(\dfrac{U_{\parallel}\,n_{\mathrm{e-h}}\,\omega}{1-U_{\parallel}\,\rho_{\mathrm{F}}}\right)^{2}} \\ 
& = \dfrac{\omega T_{\parallel}}
{(1-U_{\parallel}\,\rho_{\mathrm{F}})\,U_{\parallel}}
\,\dfrac{1}
{1+(\omega T_{\parallel})^{2}}=
\dfrac{\chi(0)}{U_{\parallel}\,\rho_{\mathrm{F}}}
\,\dfrac{\omega T_{\parallel}}
{1+(\omega T_{\parallel})^{2}},
\end{split}
\end{equation}
where in the last step we used the expression for the static spin-susceptibility
$\chi_{\parallel}(0) =\rho_{\mathrm{F}}/(1-U_{\parallel}\,\rho_{\mathrm{F}})$
and we defined 
\begin{equation}
T_{\parallel}=\dfrac{U_{\parallel}\,n_{\mathrm{e-h}}}{1-U_{\parallel}\,\rho_{\mathrm{F}}},
\end{equation}
which is the result quoted in the main text in Eq.~\ref{eq:Tlong}.

\section{Transverse relaxation within the Landau-Lifshitz-Gilbert equation}
\label{appendix:LLG}

We consider the Landau-Lifshitz-Gilbert (LLG) equation describing the damped precessional motion of a magnetic moment placed in a static external magnetic field that has been perturbed by a time-dependent transverse magnetic field:
\begin{equation}\label{eq:LLG-eom}
\frac{\ud\VEC{m}}{\ud t} = -\gamma\,\VEC{m} \times \VEC{B}^{\text{ext}} + \eta\,\frac{\VEC{m}}{m_0} \times \frac{\ud\VEC{m}}{\ud t},
\end{equation}
with $\VEC{B}^{\text{ext}} = B_0\,\hat{\VEC{z}} + \textbf{b}(t)$ and $\textbf{b}(t)=b_x(t)\,\hat{\VEC{x}}+b_y(t)\,\hat{\VEC{y}}$.
Here $B_0$ is the static part of the external field, and $\textbf{b}(t)$ a small transverse perturbing field, $|\textbf{b}(t)| \ll B_0$.
The precession rate in Eq.~\ref{eq:LLG-eom} is set by $\gamma$, while the relaxation is controlled by $\eta$, namely the damping term.
Writing $\VEC{m}(t) = m_0\,\hat{\VEC{z}} + \VEC{m}(t)$ with $\textbf{m}(t)=m_x(t)\,\hat{\VEC{x}}+m_y(t)\,\hat{\VEC{y}}$, and linearizing the LLG equation yields the following equation of motion for the transverse components of the magnetization:
\begin{align}
  \frac{\ud m_x}{\ud t} &= -\gamma\,B_0\,m_y + \gamma\,m_0\,b_y- \eta\,\frac{\ud m_y}{\ud t} ,
  \label{eq:LLG-lin-x} \\
  \frac{\ud m_y}{\ud t} &= -\gamma\,m_0\,b_x + \gamma\,B_0\,m_x + \eta\,\frac{\ud m_x}{\ud t}  
  \label{eq:LLG-lin-y} .
\end{align}

To illustrate the relaxation dynamics, consider a static external field ($\VEC{b}(t)=0$) and a small tilt of the magnetic moment at some initial time: $\textbf{m}(t=0)=m_0\left(\cos\theta\,\hat{\VEC{z}}+\sin\theta\,\hat{\VEC{x}}\right)\approx m_0\,\hat{\VEC{z}}+\Delta m\,\hat{\VEC{x}}$.
Since the expected solution is a damped precession that relaxes towards the direction of the static magnetic field, we use the following ansatz corresponding to a circular precession 
that decays in time with a transverse relaxation time $T_{\perp}$:
\begin{align}
  m_x(t) &= \Delta m\,e^{-t/T_{\perp}}\cos(\omega_0 t)  , \\
  m_y(t) &= \Delta m\,e^{-t/T_{\perp}}\sin(\omega_0 t) .
\end{align}
Plugging the above ansatz back into the LLG Eqs. \ref{eq:LLG-lin-x} and \ref{eq:LLG-lin-y} we get
\begin{equation}
  \frac{1}{T_{\perp}}\cos(\omega_0 t) + \omega_0\sin(\omega_0 t)= 
\eta\,\omega_0\cos(\omega_0 t) + \left(\gamma B_0 - \frac{\eta}{T_\perp}\right)\sin(\omega_0 t) ,
\end{equation}
\begin{equation}
  -\frac{1}{T_{\perp}}\sin(\omega_0 t) + \omega_0\cos(\omega_0 t)= 
-\eta\,\omega_0\sin(\omega_0 t) + \left(\gamma B_0 - \frac{\eta}{T_\perp}\right)\sin(\omega_0 t) .
\end{equation}
The above equations can only be satisfied if the coefficients in front of the time-dependent sines and cosines match.
We then have (both equations give the same pair of relations)
\bek
  &\dfrac{1}{T_{\perp}} = \eta\,\omega_0 ,\label{eq:LLG-Ttrans}\\
 & \omega_0 = \dfrac{\gamma B_0}{1+\eta^2} \label{eq:LLG-wmax}.
\ek
Importantly, Eq.~\ref{eq:LLG-Ttrans} shows that the transverse relaxation time is given by the product between the damping term $\eta$ and the characteristic frequency $\omega_0$.

We next turn to calculate the transverse dynamic spin-susceptibility within the LLG model.
For this, we consider the following Fourier transforms,
\begin{equation}
  \VEC{b}(t) = \int\frac{\ud\omega}{2\pi}\,e^{-\iu\omega t}\,\VEC{b}(\omega) ,\;
  \VEC{m}(t) = \int\frac{\ud\omega}{2\pi}\,e^{-\iu\omega t}\,\VEC{m}(\omega).
\end{equation}
Inserting the above expressions into the linearized equations \ref{eq:LLG-lin-x} and \ref{eq:LLG-lin-y} we obtain in frequency space
\begin{align}
  -\mathrm{i}\,\omega m_x(\omega) &= + \gamma m_0 b_y(\omega) + \left(\mathrm{i}\,\eta \omega-\gamma B_0\right) m_y(\omega) , \\
  -\mathrm{i}\,\omega m_y(\omega) &= -\gamma m_0 b_x(\omega) + \left(\gamma B_0 - \mathrm{i}\,\eta \omega\right) m_x(\omega).
\end{align}
The above can be simplified by considering the circular components 
$m_{\pm}=m_{x}\pm \mathrm{i}\,m_{y}$, yielding
\begin{equation}\label{eq:LLG-pre-chi}
\begin{split}
  \omega m_{\pm}(\omega)&=\gamma_{\pm}\big(m_{0}b_{\pm}(\omega)- B_{0}m_{\pm}(\omega)\big), \Rightarrow \\
&\Lambda_{\pm}(\omega) m_{\pm}(\omega) = b_{\pm}(\omega),
\end{split}
\end{equation}
with $b_{\pm}=b_{x}\pm \mathrm{i}\,b_{y}$, $\gamma_{\pm}=\pm \gamma/(1\mp \mathrm{i}\,\eta)$ and
\bek\label{eq:LLG-lambda}
\Lambda_{\pm}(\omega) = \dfrac{B_0}{m_{0}} + \dfrac{\omega}{\gamma_{\pm}m_{0}}.
\ek
It is apparent from Eq.~\ref{eq:LLG-pre-chi} that the transverse spin-susceptibility can be obtained from the inverse of $\Lambda_{\pm}(\omega)$ defined above.
After some algebra and picking the minus sign in Eq.~\ref{eq:LLG-lambda} one obtains
\begin{equation}
\begin{split}
\chi^{\text{LLG}}_{\pm}(\omega)&=\big(\Lambda_{-}(\omega)\big)^{-1}  
=\dfrac{m_{0}\omega_{0}}{B_{0}}
\cdot\dfrac{-\omega+(1+\eta^{2})\omega_{0}+\mathrm{i}\,\eta\omega}{(\omega-\omega_{0})^{2}+(\eta\omega_{0})^{2}}.
\end{split}
\end{equation}
The density of spin-excitations in the LLG model are thus described by a skewed Lorentzian in $\omega$:
\bek\label{eq:LLG-imchi}
\Im \chi^{\text{LLG}}_{\pm}(\omega)=
\dfrac{m_{0}\omega_{0}}{B_{0}}
\dfrac{\eta\omega}{(\omega-\omega_{0})^{2}+(\eta\omega_{0})^{2}}.
\ek
The resonance frequency of the above function takes place at
\bek
\dfrac{\text{d}}{\text{d}\omega}\Im \chi^{\text{LLG}}_{\pm}(\omega)=0\Rightarrow
\omega_{\text{res}}=\sqrt{1+\eta^{2}}\omega_{0},
\ek
while the FWHM amounts to
\bek\label{eq:Gamma-eta-w0}
\Gamma= 2\eta\omega_{0}\label{eq:LLG-FWHM}
\dfrac{\sqrt{2+3\eta^{2}+2\sqrt{1+\eta^{2}}}}{1+\sqrt{1+\eta^{2}}}
\simeq 2\eta\omega_{0}.
\ek
We note that the above approximation is exact in the 
$\eta\rightarrow 0$ limit and involves only a $\sim10\%$ relative error for
$\eta=1$, which is by far the maximum value that damping can get for 
single adatoms;
for most of the elements analyzed in the main text we have 
$\eta\lesssim 0.5$~\cite{ibanez-azpiroz_zero-point_2016}, 
so the approximation of Eq.~\ref{eq:LLG-FWHM}
is indeed very good. Then, comparing Eq.~\ref{eq:LLG-FWHM} to 
Eq.~\ref{eq:LLG-Ttrans} we arrive to the relation between the FWHM and the transverse relaxation time
quoted in the main text:
\bek
T_{\perp}=\dfrac{2}{\Gamma}.
\ek


\begin{thebibliography}{100}
\bibitem{footnote}
For externally applied $B= 517$ T the magnetic moment of Ni is $m=0.29$
  $\mu_{B}$, while a Fe adatom placed as a second nearest neighbor induces a
  magnetic moment on Ni of $m=0.51$ $\mu_{B}$.
\end{thebibliography}

\end{document}